\documentclass[12pt]{iopart}

\usepackage{graphicx}
\usepackage[table]{xcolor} 
\usepackage{booktabs} 
\usepackage{upgreek}
\usepackage{caption}

\begin{document}
\title{Automated Optical Inspection and Image Analysis of Superconducting Radio-Frequency Cavities}
\author{Marc Wenskat}
\address{Deutsches Elektronen Synchrotron, 22607 Hamburg, Notkestrasse 85}
\ead{marc.wenskat@desy.de}
\vspace{10pt}
\begin{indented}
\item[]March 2017
\end{indented}

\begin{abstract}
The inner surface of superconducting cavities plays a crucial role to achieve highest accelerating fields and low losses. For an investigation of this inner surface of more than 100 cavities within the cavity fabrication for the European XFEL and the ILC HiGrade Research Project, an optical inspection robot OBACHT was constructed. To analyze up to 2325 images per cavity, an image processing and analysis code was developed and new variables to describe the cavity surface were obtained. The accuracy of this code is up to 97\,\% and the PPV 99\,\% within the resolution of 15.63 $\upmu \mathrm{m}$. The optical obtained surface roughness is in agreement with standard profilometric methods. The image analysis algorithm identified and quantified vendor specific fabrication properties as the electron beam welding speed and the different surface roughness due to the different chemical treatments. In addition, a correlation of $\uprho = -0.93$ with a significance of $6\,\upsigma$ between an obtained surface variable and the maximal accelerating field was found. 
\end{abstract}

\section{Introduction}
Superconducting niobium radio-frequency (RF) cavities are fundamental for accelerators like the European X-Ray Free Electron Laser (XFEL), the International Linear Collider or LCLS-II \cite{XFEL_TDR,ILC_TDR,LCLS-II}. Electromagnetic RF fields, used for the particle acceleration, penetrate the inner surface of the resonator volume. For an investigation of the behavior of cavities under such RF fields, an optical surface inspection was developed at KEK and Kyoto University, the "Kyoto camera system" \cite{Iwashita2008,Tajima2008}. The intention is, that an optical inspection of the inner surface, which is exposed to the RF field, leads to a better understanding of limitations observed during RF test. While a general correlation was found between low field quenches (transition from superconducting to normal-conducting phase) and localized defects seen in optical inspections \cite{Watanabe,Moller2009,Geng2009a,Aderhold2010b,Singer2010}, the interplay of global surface properties and RF performance still needs to be investigated. A method to study possible relations between the surface properties and the RF performance is presented in this paper.

\section{Optical Inspection Robot}
\begin{figure}[htbp]
	\centering
		\includegraphics[width=0.7\textwidth]{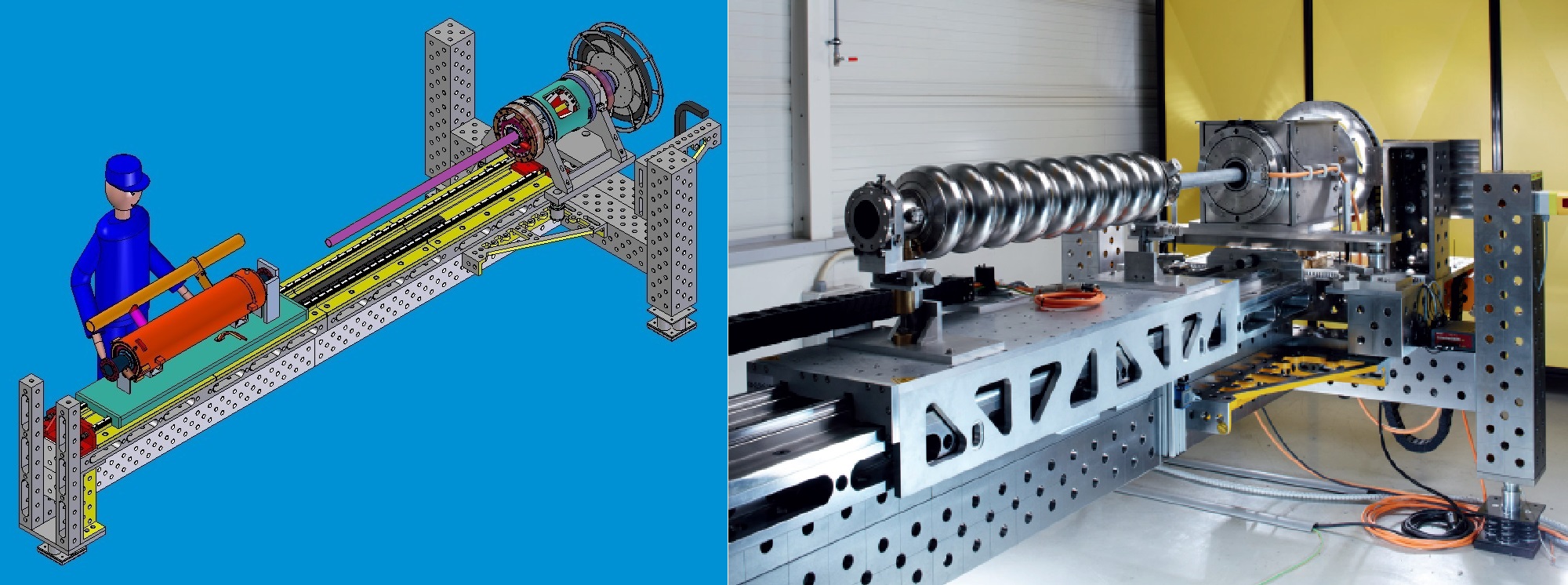}
	\caption{3D sketch of the OBACHT set up on the left. A cavity with helium tank (orange) is mounted on the movable sled (turquoise), which is located at the mounting position. The camera tube (purple) can be rotated with the torque motor (turquoise/gray). A photography of the existing setup with an undressed cavity mounted on the sled is shown on the right.}
	\label{fig:Anlage_komplett_ISO_II_cropped}
\end{figure}

A first setup of the Kyoto camera system was implemented at DESY, where optical inspections of pre-series European XFEL cavities and other cavity test series were performed \cite{Sebastian}. The sliding table, the camera rotation and the data handling were completely manually operated. A single inspection took up to two days until it was finished. The examination of the images were partially done while they were taken by the operator, but in general took an additional day. \newline 
To establish the optical inspection as a quality management tool during the production of 832 SRF cavities for XFEL, an automated optical inspection tool with automatic data management was developed at DESY as a successor of the original Kyoto camera setup. The aim of this project was to reduce the total time of an optical inspection as well as the time which an operator has to actively monitor the process. This was vital since a tool for qualification management needs short times between receiving a cavity and reporting a feedback to the production line. 
The result of the above mentioned development is the "\textbf{O}ptical \textbf{B}ench for \textbf{A}utomated \textbf{C}avity inspection with \textbf{H}igh resolution on short \textbf{T}imescales" (OBACHT) and has already proven its potential as a tool for quality control for the vendors producing cavities for XFEL. In Figure \ref{fig:Anlage_komplett_ISO_II_cropped}, a 3D sketch of OBACHT and a photography is shown.
\newline
The 9-cell-cavity is mounted on a movable sled. The linear motor, which drives the movable sled, has a positioning precision of $\pm 2\,\upmu \mathrm{m}$. The torque motor for the rotation of the camera tube has a positioning precision of $\pm 0.04^0$. A more detailed description of all mechanical and sensor components can be found in \cite{Lemke,Wenskat2015}.

\subsection{Optical System}
\begin{figure}[htbp]
	\centering
		\includegraphics[width=0.75\textwidth]{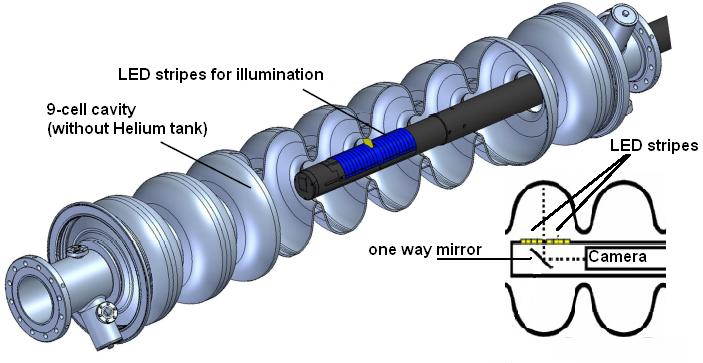}
	\caption{Drawing of the Kyoto Camera System used at DESY. The camera is viewing the inner surface via a $45^\mathrm{o}$ tilted one way mirror. The distance between the camera and the mirror can be controlled for focusing. Behind this one way mirror, three LEDs are mounted for the central illumination. $2 \times 10$ acrylic strips with LEDs are mounted left and right from the opening in the tube for a more detailed illumination \cite{Lemke}. }
	\label{fig:sketch}
\end{figure}

\begin{figure}[htbp]
	\centering
		\includegraphics[width=0.4\textwidth]{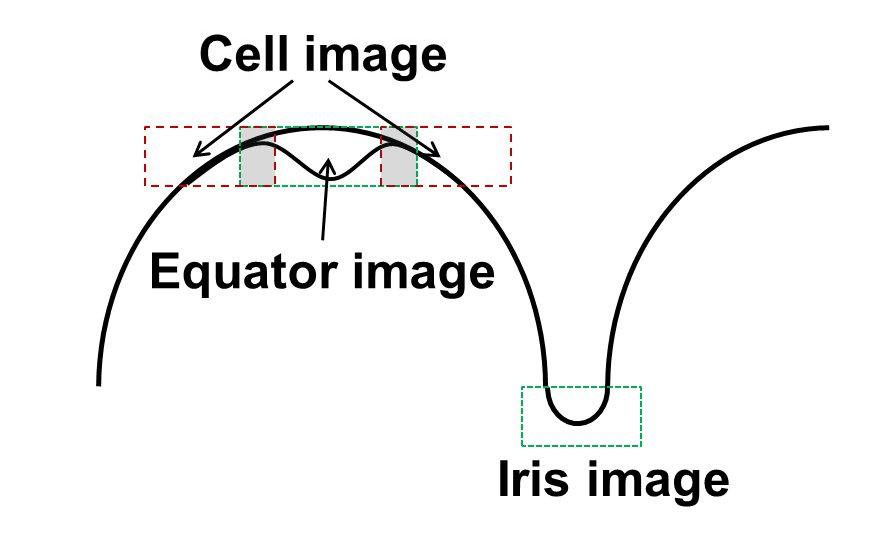}
	\caption{Image positions in a cavity (not to scale). At the equators, the welding seam itself is in the central axis of the image - depicted by the green box. The cell images are taken with an offset of $\pm 8\,\mathrm{mm}$ relative to the equator position - depicted by the red boxes. The overlap region is shaded. The iris image is taken between the cells and at the beam pipe welding.}
	\label{fig:Image_Positions}
\end{figure}

The camera system developed at KEK and Kyoto in 2008 \cite{Iwashita2008,Tajima2008} was implemented at DESY \cite{Sebastian} and had a major upgrade in 2009 \cite{Sebastian,Iwashita2009}. An overview of the camera system is given in Figure \ref{fig:sketch}. 
\newline
It consists of a camera tube with a diameter of 50\,mm to fit into the cavity without colliding with the irides or antennas protruding into the cavity volume. In this tube, the camera together with a low-distortion lens (LM75JC by Kowa Industrial) are installed. The camera system images the surface via a $45^\mathrm{o}$ tilted one way mirror which can be continuously adjusted to other angles in order to inspect any cavity region. The focal distance of the camera to the cavity surface is controlled by a motor driven lead-screw. For illumination, acrylic strips (two LEDs per strip) are attached to the camera tube around the camera opening, together with three additional LEDs behind the one way mirror inside the camera tube.
\newline
With given cavity geometry and optical set up, an individual image covers $5^\mathrm{o}$ of an equator as well of the cell image. To have a small overlap at the edges of an image, an image is taken each $4.8^\mathrm{o}$. This yields to 75 images per equator and 675 equator images per cavity and two times 675 cell images per equator, while the iris images are taken with an angular spacing of $12^\mathrm{o}$ and yield to 30 images per iris. Additionally, there is an overlap along the cavity z-axis for the cell and equator images, see Figure \ref{fig:Image_Positions}.

Each longitudinal position, equator, cell and iris, has an optimized illumination pattern, accounting for the individual surface geometry and reflectivity in the respective region, as well as individual focus settings. With given angular and longitudinal spacing, a total of 2325 images per automated inspection are taken. If needed, individual images of the cavity surface with manual controllable settings can also be taken. The highest magnetic field in a cavity, and hence the highest losses, are at the equatorial welding seam region including the heat affected zone. Therefore, the equatorial images are of main interest for this analysis. The image format used at present is JPEG, which is a widely used image compression format and the average size of such an image is on the order of 5\,MB.

\subsubsection{Camera}
\label{section:OSDOF}
Within the tube, a digital camera (Artray MI900) is installed and connected via USB to the PC. The camera sensor is a CMOS Bayer pattern with $3488 \times 2616$ pixels on a sensor area of $6.17 \times 4.55\,\mathrm{mm}^2$. It has a theoretical Signal-to-Noise-Ratio (SNR) of 35 dB. An investigation with a noise estimator \cite{Liu2013} at OBACHT showed that the effective SNR is 32\,dB. The primary source of noise in this imaging system is shot noise, which is an intrinsic property of the CMOS sensor.

Important properties of an optical system are the resolution and the depth of field (DOF). Theoretical calculations in \cite{Sara1995,Xie2009} have shown that accelerating fields of 50\,MV/m are achievable if surface structures and localized defects are below 10\,$\upmu \mathrm{m}$, hence the resolution of the system should be on that order. The optical system is a diffraction limited lens system which further is projected onto a digital camera sensor. A rough estimation of the resolution $d$ of the lens system can be done via the Abbe diffraction limit. The numerical aperture $NA$ of the system is 0.4, the wavelength of optical light varies between 400 - 800\,nm. This yields to a resolution of the lens system of $d = 1-2\,\upmu \mathrm{m}$. The digital camera further decreases the resolution of the whole optical system. 

A more precises calculation of the theoretical limit of the resolution of the camera-lens system can be deduced with the Point Spread Function (PSF) of the system. The PSF describes the behavior of the light, emitted from a point-like light source, as it is transferred through the optical system and finally detected at the imaging plane. The calculation of the theoretical PSF was done with a Java implementation of a PSF Generator \cite{Sage2014,Griffa2010} and MATLAB to control and analyze the process. The Born-Wolf Model \cite{Born2003,Frisken-Gibson1989} was chosen as suitable for the setup since this model describes the diffraction of a spherical wave by a circular aperture when the point light source is in focus and no immersion oils are used. The minimum distance which is needed to resolve two distinct objects is given via the Full-Width-Half-Maximum (FWHM) of the PSF, if the detector is in the imaging plane. With the values of the system a theoretical resolution of OBACHT of $d =  11.7 \pm 0.9\,\upmu \mathrm{m}$ can be achieved, if the object is in focus. 

The effective resolution of OBACHT was investigated using a dedicated test pattern, namely the USAF 1951 Resolution Test Chart. It is a set of well defined separate lines with decreasing distance and thickness. The smallest distance between objects, which were still resolved as an array of separate lines, is the effective resolution of the system. For OBACHT, these were the elements of group number five, element number three, which is equivalent to an effective resolution of $d =  12.4\,\upmu \mathrm{m}$. This is in agreement with the theoretical limit, since the next finer group of elements is below the theoretical resolution.
\newline
The magnification M can be calculated from the given projected cavity surface seen in the image ( $12 \times 9\,\mathrm{mm^2}$) and sensor dimensions ($6.1 \times 4.5\,\mathrm{mm}^2$) and yields to M = 0.51, since the image is larger than the sensor. With given magnification M and object resolution, the camera resolution can be calculated to be 6.31\,$\upmu \mathrm{m}$. With the given pixel sensor size of 1.75\,$\upmu \mathrm{m}$, the smallest resolvable objects have a size of four pixels. This is in agreement with the theoretical resolution, where the maximum of the PSF distribution covers four pixels.

\begin{figure}[htbp]
	\centering
		\includegraphics[width=0.75\textwidth]{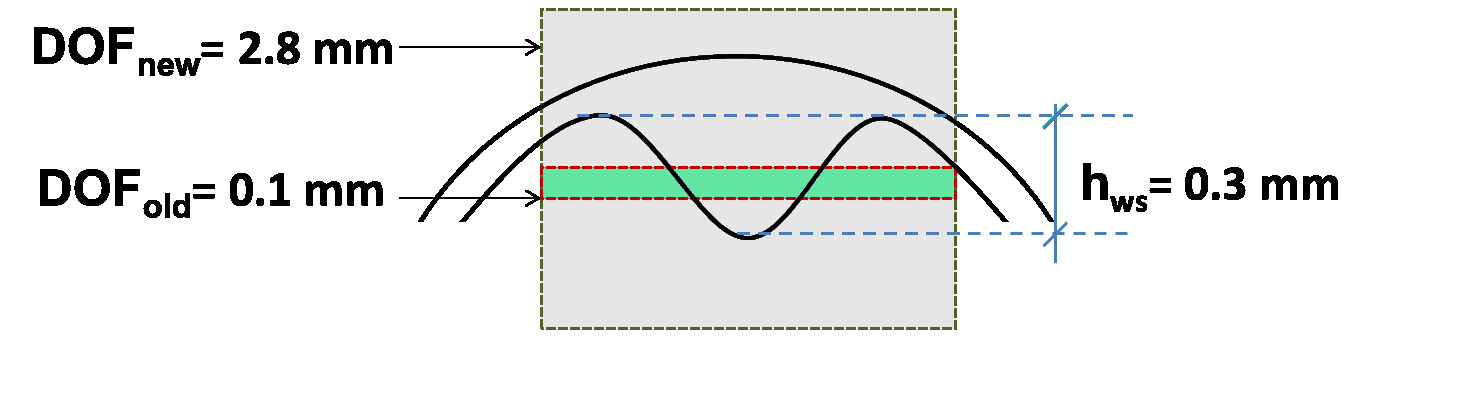}
	\caption{A schematic cross section of the welding seam surface. The old DOF was less than one third of the welding seam height and enabled the system to perform height map scans but led to blurry regions. The new DOF covers the whole range of the surface profile and each part of the image is in focus.}
	\label{fig:DOF_1}
\end{figure}

\begin{figure}[htbp]
	\centering
		\includegraphics[width=0.3\textwidth]{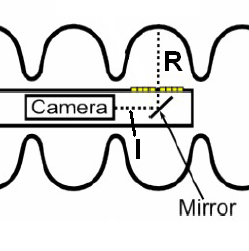}
	\caption{The focal length of the optical system $f$ is the sum of the cell radius R and the camera-mirror distance l. The latter has to be adjusted to correct for deviations of the surface-mirror distance to keep the inner cavity surface in focus.}
	\label{fig:Bildweite}
\end{figure}

\begin{figure}[htbp]
	\centering
		\includegraphics[width=0.5\textwidth]{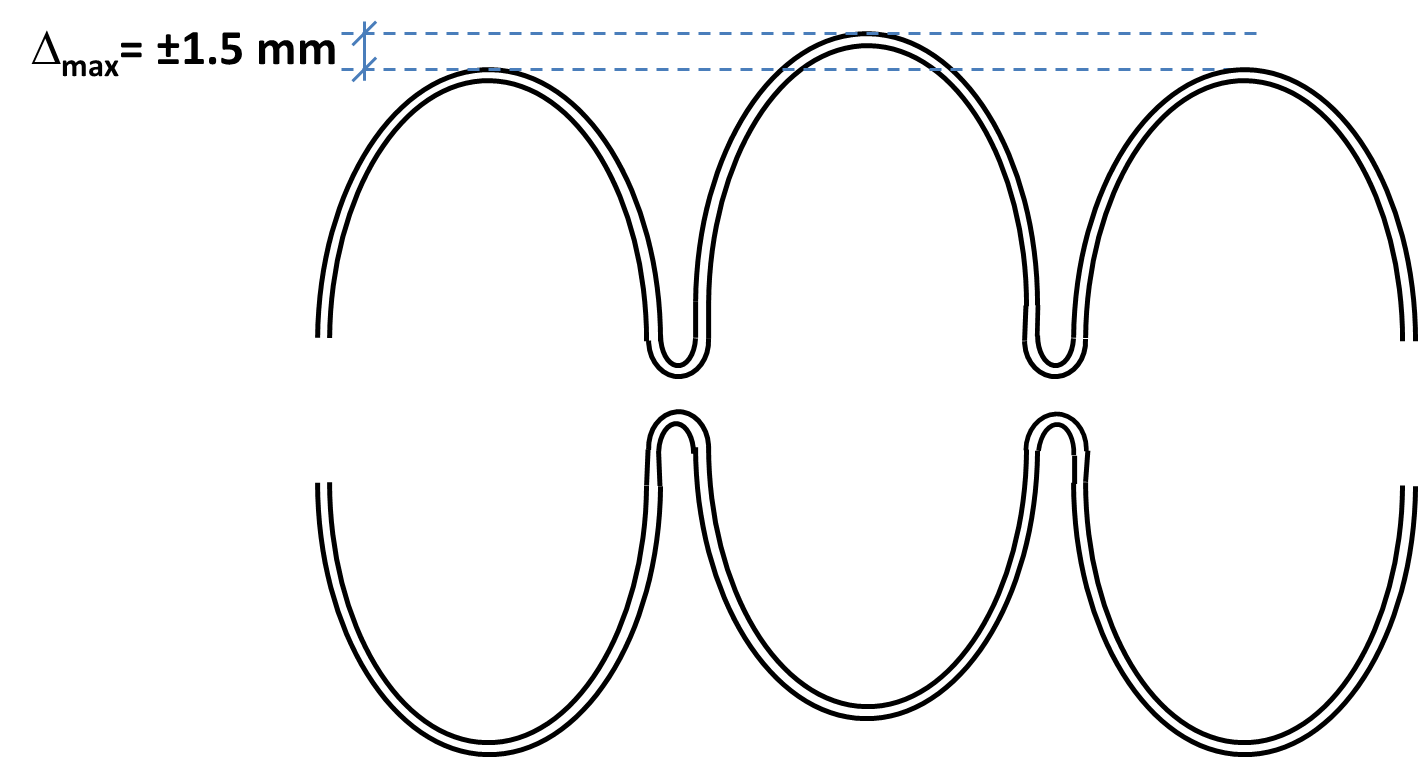}
	\caption{A cell symmetry axis can deviate from the geometrical axis up to 1.5\,mm. The deviations are randomly angular distributed and each cell can be affected, leading to a individual camera position for each cell.}
	\label{fig:DOF_2}
\end{figure}

The DOF is defined as the distance between the nearest and farthest objects in an image, which are still in focus. In general, the DOF is controlled with the F-number of the lens, because the DOF is inverse proportional to the relative aperture. The aperture is set manually at the lens, and a disassemble of the camera tube is necessary to change it. The original DOF was less than 0.1\,mm \cite{Sebastian}, but experience showed that a DOF of 2.8 mm is necessary for an automated optical inspection in order to achieve focused images. The equatorial welding seam region has a W-like cross section, see Figure \ref{fig:DOF_1}. The welding seam itself rises up to 0.3 mm above the surrounding cavity surface. Although a small DOF gives the possibility to make a profile scan along the lateral axis and create a height map of the welding seam via focal stacking \cite{FS1,Forster2004}, it would also increase the amount of unfocused regions in a single image and decrease the spatial resolution in these regions. 
\newline
To control the center plane of the DOF, the camera focus, the camera is moved with a lead screw in the camera tube, see Figure \ref{fig:Bildweite}. 
It is driven by a torque motor and has a precision of 4\,$\upmu \mathrm{m}$. The optical system has a focal length of f = 150\,mm and the surface-mirror distance equals the cell radius R and is in the order of 103 $\pm$ 0.3 mm. The offset of 0.3 mm is found to be the average deviation for XFEL production  and is measured w.r.t. to the cavity geometrical axis defined by the centers of two reference rings \cite{Gresele2013,Sulimov} but cell-to-cell deviations can be up to 1.5\,mm, see Figure \ref{fig:DOF_2}. These deviations are the cause for an optical calibration. 
\newline
The optimal camera position for a specific cell can be deduced by finding the best camera position at four different angles around the cell, each $90^\mathrm{o}$ apart, and calculate the average position in which each image at the four angles is in focus. This autofocus algorithm is realized by moving the camera position within a predefined range and certain number of steps from a starting point, then taking an image at every step. The images are analyzed and the camera position with the best focus is determined, based on an algorithm described in \cite{Raquel,Autofocus1,Autofocus2,Toropov} This procedure is repeated for each cell. With these positions along the cavity, a linear fit, the optical axis, is derived. The camera can now be set to the best position for each cell deduced from this optical axis. The DOF of 2.8\,mm reduces the range of adjustment of the camera position to compensate offsets to the optical axis.

\subsubsection{Illumination System}
\label{sec:licht}
\begin{figure}[htbp]
	\centering
		\includegraphics[width=0.5\textwidth]{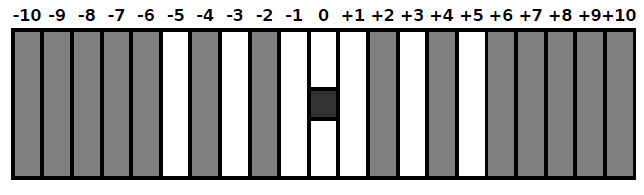}
	\caption{Default illumination pattern for images at the equator \cite{Sebastian}. Stripes are numbered from left to right, starting with -10 and ending with +10. Grey stripes are switched off, white stripes are switched on. The dark square in the center indicates the opening for the camera.}
	\label{fig:Illumination_Pattern}
\end{figure}

\begin{figure}[htbp]
	\centering
		\includegraphics[width=0.5\textwidth]{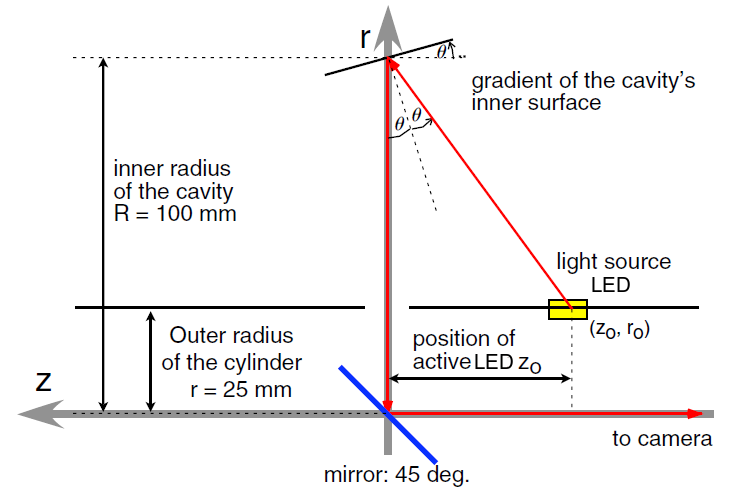}
	\caption{Schematic view of the illumination model \cite{Iwashita2008}. With given properties of the cavity and acquisition geometry, the incident angle $\theta$ of a grain boundary can be calculated.}
	\label{fig:Illumination_Geometry}
\end{figure}

The purpose of the illumination system is to guarantee an intensity distribution over the image which enhances surface structures like grain boundaries but minimizes shadows. At OBACHT, three LEDs are installed behind the one way mirror, together with $2 \times 10$ acrylic strips, with two LEDs per strip, left and right of the tube opening for the camera. The acrylic strips have a width of 7 mm and can be individually turned on and off. The three LEDs behind the one way mirror are used to compensate the missing LED strip at the camera opening.  
\newline
The capability to adapt the image illumination system to the surface geometry in order to achieve the optimized intensity distribution over the image is limited. Due to the cross section of the welding seam, it is unavoidable to have shadows and illumination pattern introduced by the LED strips in the image. The default setting of the illumination pattern at OBACHT, shown in Figure \ref{fig:Illumination_Pattern}, was based on experience and later reevaluated with ray-tracing methods \cite{Steel}. 
The detection of a given surface feature at an angle $\theta$ to the nominal surface plane is determine by the geometry shown in Figure \ref{fig:Illumination_Geometry}. This incident angle depends on the slope of the boundary and the distance to the LED stripe and the camera. With the given geometry, a maximum boundary slope of $20^\mathrm{o}$ can be resolved. This value was also derived in \cite{Watanabe2008}. Any boundary with a slope above $20^\mathrm{o}$ will be darker than its surroundings and is still detected, but an determination of the boundary slope is impossible.

\section{Image Processing}
OBACHT provides visible information on the cavity surface structure which includes grain boundaries, defects and the welding seam. These structures have to be characterized in terms of geometrical properties and location within the image by means of the analysis algorithm. To improve the detectability of the grain boundaries, the image has to be processed. A combination of global and local operators acting on the image was developed in the scope of this thesis, where the goal is to enhance features in the image. A flowchart of the processing steps is shown in \ref{fig:Flowchart}.

\subsection{Digital Image and Objects of Interest}
\label{sec:DigImg}
The presentation of the digital image taken by OBACHT, as used as an input to the algorithm, is a $3488 \times 2616 \times 3$ matrix. The first two dimensions represent the spatial dimension of the image. The third dimension represents the color information in the image. The three color layers are red (R), green (G) and blue (B). The objects of interest are the grain boundaries of the niobium crystal. An optical boundary in an image is defined as a contour with finite width of up to several pixels that represents a continuous change of intensity. In contrast, an edge is the border of a boundary. In \cite{Martin2004} it is stated, that "edge detection can be a low-level technique towards the goal of boundary detection". Those two definitions, that of an edge and that of a boundary, should not be confused. The physical grain boundary is used synonymously with the optical boundary in this paper. The motivation for this is that any intensity gradient, which is in an image boundary, can only be caused by a geometric gradient or by a change in reflectivity. The first is a grain boundary, the latter can be caused by impurities or different surface structures.

\subsection{Processing Algorithm}
\begin{figure}[htbp]
	\centering
		\includegraphics[width=0.4\textwidth]{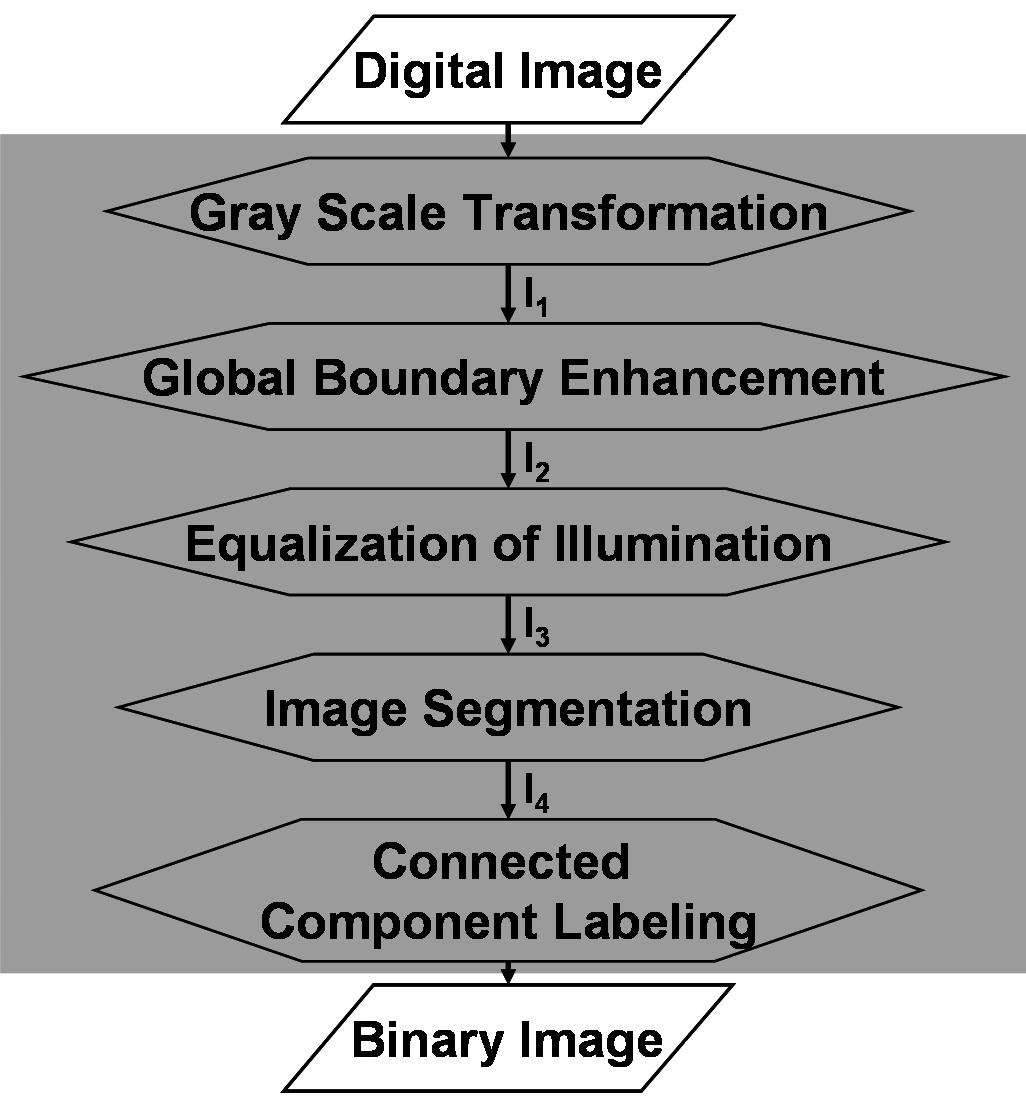}
	\caption{Flowchart of the developed algorithm for image processing. The output of the algorithm is a binary image. The image processing algorithm is depicted by the gray rectangle. The indices '$I_i$' refer to intermediate pictures. Both, the binary image and the original digital image is used for further image analysis.}
	\label{fig:Flowchart}
\end{figure}
\subsubsection{Gray scale transformation}
Since the physical objects of interest are the grain boundaries, the objects of interest for the image processing algorithm are the optical boundaries. To simplify the operators used in the algorithm, the color image is transformed into a gray scale image. Mathematically, the image matrix is reduced from three color to one intensity dimension. To achieve this, a weighted sum of the color components for each pixel is calculated via
\begin{equation}
I_1 = 0.2989 \cdot R + 0.5870 \cdot G + 0.1140 \cdot B  
\end{equation}
with $R$,$G$ and $B$ stand for the respective color layer and $I_1$ the intensity for the pixel. The weights are taken from the CIE1931 standard color space transformations \cite{COL:COL20020}. This transformation causes no loss of relevant information although the color is neglected since the intensity information is conserved during this transformation which carries all information about relevant gradients and connectivity of pixel. The color on the other hand can be a powerful descriptor that often simplifies object identification and is preserved in the original image and available later for further image analysis steps. 

\subsubsection{Global boundary enhancement}
As a second step, all pixels with intensity gradients will be enhanced, regardless of the origin of their gradients. This is done by generating an intermediate image which is obtained after applying a Gaussian low-pass filter to the gray scale image $\mathrm{I_{1}}$. This intermediate image is inverted and added to the gray scale image $\mathrm{I_{1}}$. Since the intermediate image will only contain low spatial frequency parts and the intensity values are inverted, the intensity values of pixel with low spatial frequency parts will be reduced. The resulting image $\mathrm{I_{2}}$ will contain pixels with high gradients and high intensities and pixels with low gradients and low intensities. This step simplifies the further detection of boundaries.
\newline
Using a Gaussian filter before boundary detection reduces the noise level in the image by smoothing the fluctuations, which improves the result of the following boundary detection algorithm. The kernel of the low pass filter is a $n_x \times n_y$ = 5 $\times$ 5 pixels matrix with a standard deviation $\upsigma$ of five. 
The value for $\upsigma$ and the kernel was chosen to be suitable with the given OBACHT resolution. If $\upsigma$ and the kernel size would be larger, the overlapping of neighboring objects would increase and the topology could be changed. A smaller value of the kernel would not be enough to cover noise objects, since OBACHT has a resolution in the order of four pixel. A smaller value of $\upsigma$ would create a steeper Gaussian filter and the effectiveness of the noise reduction would decrease. Hence, the smallest odd integer which fulfill these criteria is five.  
\newline
The resulting Gaussian filter is convoluted with $\mathrm{I_{2}}$. Each pixels new value is set to a weighted average of that pixels neighborhood. The original pixel value receives the heaviest weight (having the highest Gaussian value) and neighboring pixels receive smaller weights as their distance to the original pixel increases. Since the neighborhood of each pixel is used in this filter, the boundaries and edges are preserved better than other, more uniform blurring filters. Furthermore, it does not introduce ringing effects into the image.

\subsubsection{Equalization of Illumination}
\label{sec:secCLAHE}
The image processing algorithm has to even out the illumination pattern in order to detect boundaries regardless their position in the image. To achieve this, a local approach has to be chosen. Not the complete image but rather a local pixel neighborhood should be processed to even out the differences and hence to improve the detection probability. The so called contrast limited adaptive histogram equalization (CLAHE) algorithm is used, which in addition prevents an over-amplification of noise \cite{Zuiderveld1994, Pizer1987}. 
\newline
For further reduction of noise in the resulting image after the CLAHE algorithm, a median filter for smoothing is applied, where the idea of a median filter is to replace each pixel value with the median of its local neighborhood. This is a non-linear digital filter which preserves edges better and is less sensitive to outliers than linear filters at low noise levels \cite{Arias-Castro2009}. 

\subsubsection{Image Segmentation}
Image segmentation is a partitioning process that divides an image into regions. The method used here is a histogram based segmentation, called Otsu's method \cite{Otsu1979}. The underlying assumption for this method is that the image consists of two pixel classes, class 0 are background pixels and class 1 are foreground pixels, and the intensity histogram is a bimodal distribution. In this case the foreground objects are the boundaries and the background homogeneous regions in the image. This bimodal structure of the histogram is enhanced during the "Global Boundary Enhancement" step. Otsu's method searches for an optimal threshold t which separates the classes and minimizes the intra-class variance and maximizes the inter-class variance. 
All pixels with gray values above this threshold are considered to be foreground. The output of this algorithm is a binary image of the same size as the input image. Every foreground pixel will have the value 1 and the background pixels have the value 0. 

\subsubsection{Connected Component Labeling}
The final processing step is the so called connected component labeling. The aim of this step is to decide which pixels are connected and form a single object, like a grain boundary. The method used here is the run-length encoding \cite{Cherry67}. A row (or column) in a binary image can be represented as a sequence of ones and zeros, where a connected sequence of ones is called a run. Each single run can be represented by the position of the starting pixel and by the number of pixels in this run, which is called run-length. Based on the multiple runs in a binary image, an adjacency matrix is created, in which the spatial connections in rows and columns between several runs is encoded. A graph theoretical approach to this adjacency matrix \cite{dulmage1958,Hall01011935} is used to compute the connected components of the corresponding graph and labels each run in the output label matrix according to the adjacency matrix. Runs with same labels are considered as connected components and the output label matrix is a binary image. This binary image is the key for the image analysis algorithm since it contains the information which pixel is part of a grain boundary.

\subsection{Pixel Noise Reduction}
\label{sec:nose_cut}
\begin{figure}[htbp]
	\centering
		\includegraphics[width=0.9\textwidth]{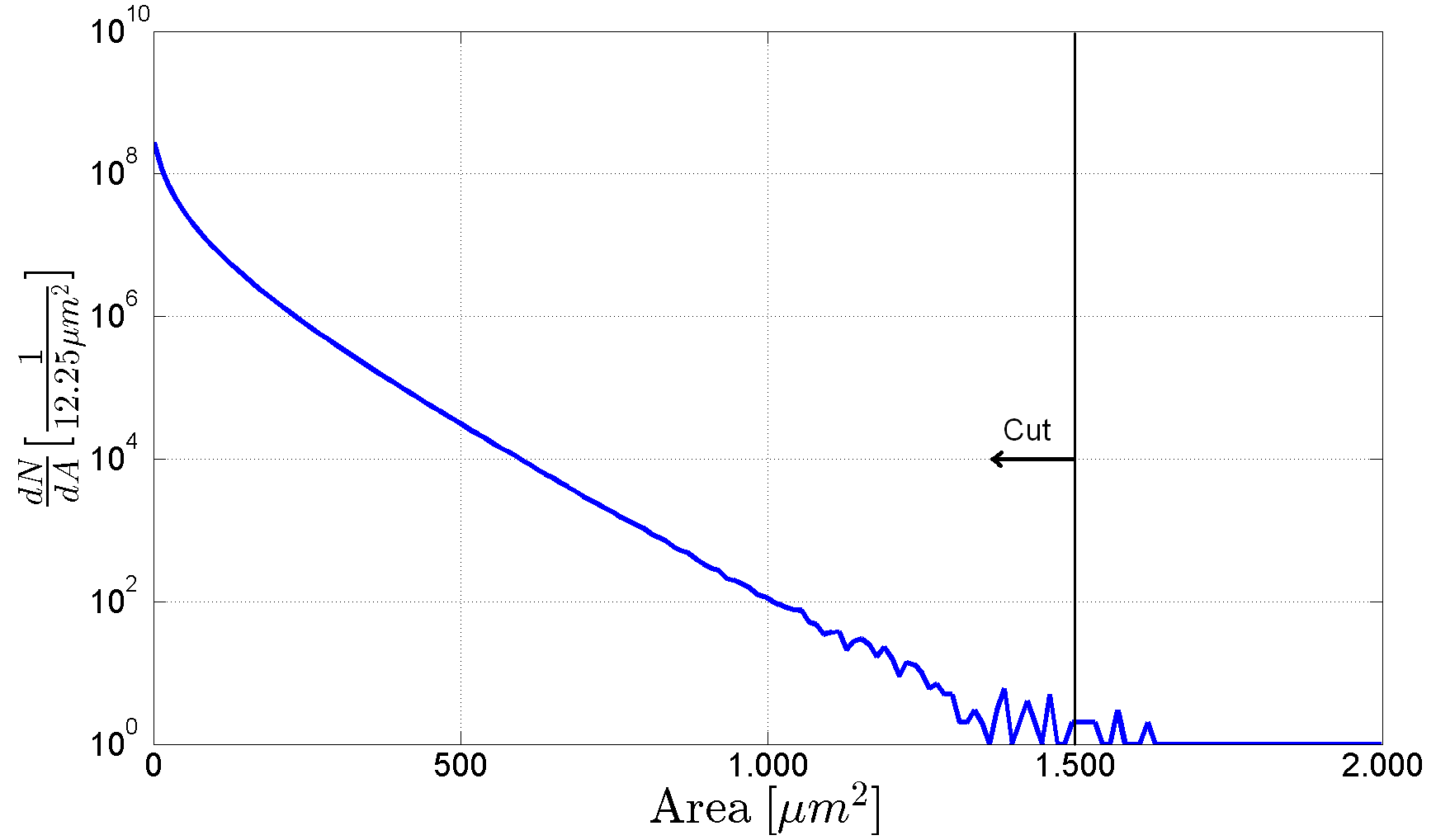}
	\caption{Pixel noise distribution. On the x-axis the area of the objects found in the binary images is shown. On the y-axis the counts per bin are displayed on logarithmic scale. This histogram contains the sum of 1000 images. Objects below an area of 1500\,$\upmu \mathrm{m^2}$ are cut.}
	\label{fig:dNdA_simulation}
\end{figure}

\begin{figure}[htbp]
	\centering
		\includegraphics[width=0.5\textwidth]{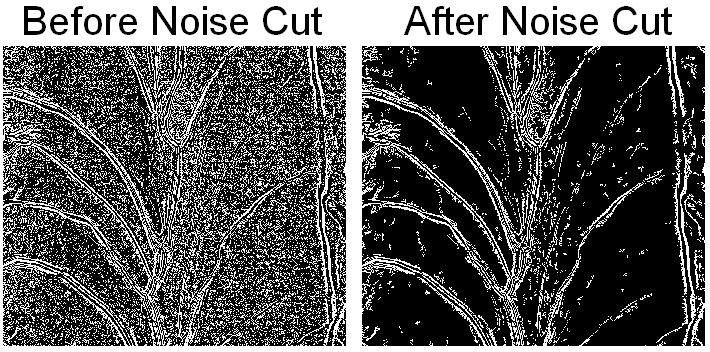}
	\caption{The left image shows an image section of the binary representation before the area cut to remove the noise, the right image the same image section after objects with an area smaller than 1500\,$\mathrm{\upmu m^2}$ are removed. More than 99.8~\% of objects in the image was removed.}
	\label{fig:NoiseCut_Comparea}
\end{figure}

The final binary image will contain unphysical objects due to shot noise fluctuations which were enhanced during the image processing, although steps were done to minimize this effect. The following possible consequences of pixel noise must be considered:
\begin{enumerate}
	\item \textit{Unphysical object generation}: If single or clusters of pixels exceed a certain intensity difference w.r.t. its local neighborhood solely because of shot noise, they are enhanced throughout the processing and would be identified as surface features. 
	\item \textit{Topological errors}: In regions with high object densities and small object distances, it is possible that two distinct objects are connected to one single object because of shot noise. 
	\item \textit{Classification errors}: Intensities along boundaries of objects vary because of the shot noise. This would lead to misclassified individual pixels, if they falsely undershoot a threshold. 
\end{enumerate}
An approach to remove these unphysical objects from (1) is shown next. The problems raised in (2) and (3) are discussed in section \ref{sec:benchmark}. 
\newline
Starting with a black $2616 \times 3488$ image, a fraction of white pixels was randomly inserted. This fraction was observed to be $25 \pm 2\%$ as the average amount of white pixels in a typical processed OBACHT image after image processing and is used as an input for this method. Although this fraction overestimates the total amount of noise since it contains real grain boundaries as well, it gives a reasonable starting point for the following procedure. 
\newline
After generating 1000 images with the random distributed white pixels and applying the image processing algorithm, the accumulated area distribution of the detected objects is derived, see Figure \ref{fig:dNdA_simulation}.  
Given the area distribution, a threshold value of 1500\,$\upmu \mathrm{m^2}$ or 122 pixels was identified by choosing the largest artificial created object by shot noise. Allmost all objects generated by shot noise are smaller than this value. A comparison of an image section before and after the area cut on this threshold value is shown in Figure \ref{fig:NoiseCut_Comparea}.
Some small remains of noisy objects can be seen in Figure \ref{fig:NoiseCut_Comparea} on the right, but more than 99.8~\% of the objects in an image are removed on average. This reduces the number of objects, which need to be analyzed to the order of several hundreds but will limit the resolution of the algorithm to a level below the optical system.

\subsection{Result of the Processing Algorithm}
\begin{figure}[htbp]
	\centering
		\includegraphics[width=0.5\textwidth]{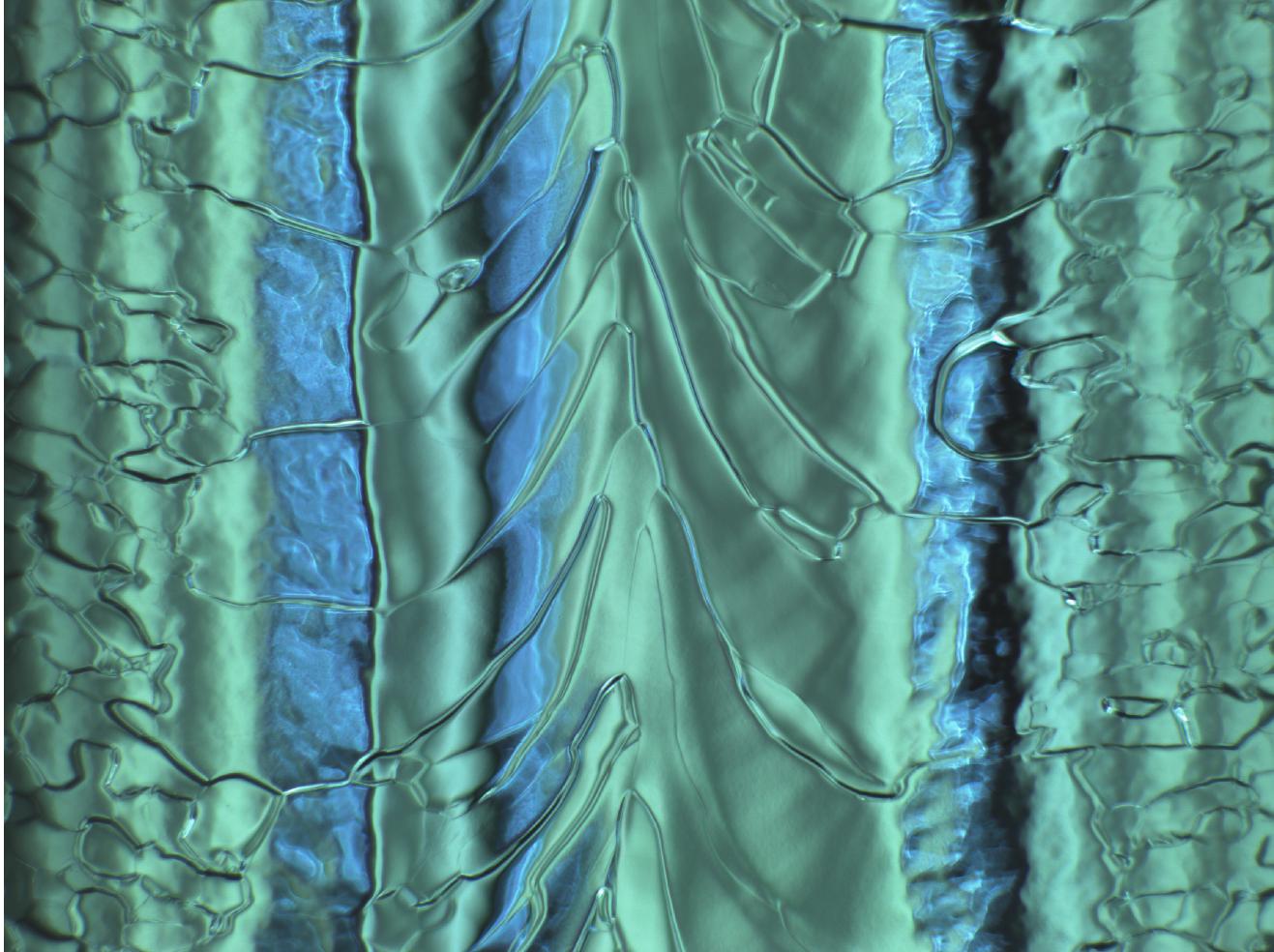}
	\caption{Digital image of a treated cavity as taken with OBACHT.}
	\label{fig:acorigina}
\end{figure}

\begin{figure}[htbp]
	\centering
		\includegraphics[width=0.5\textwidth]{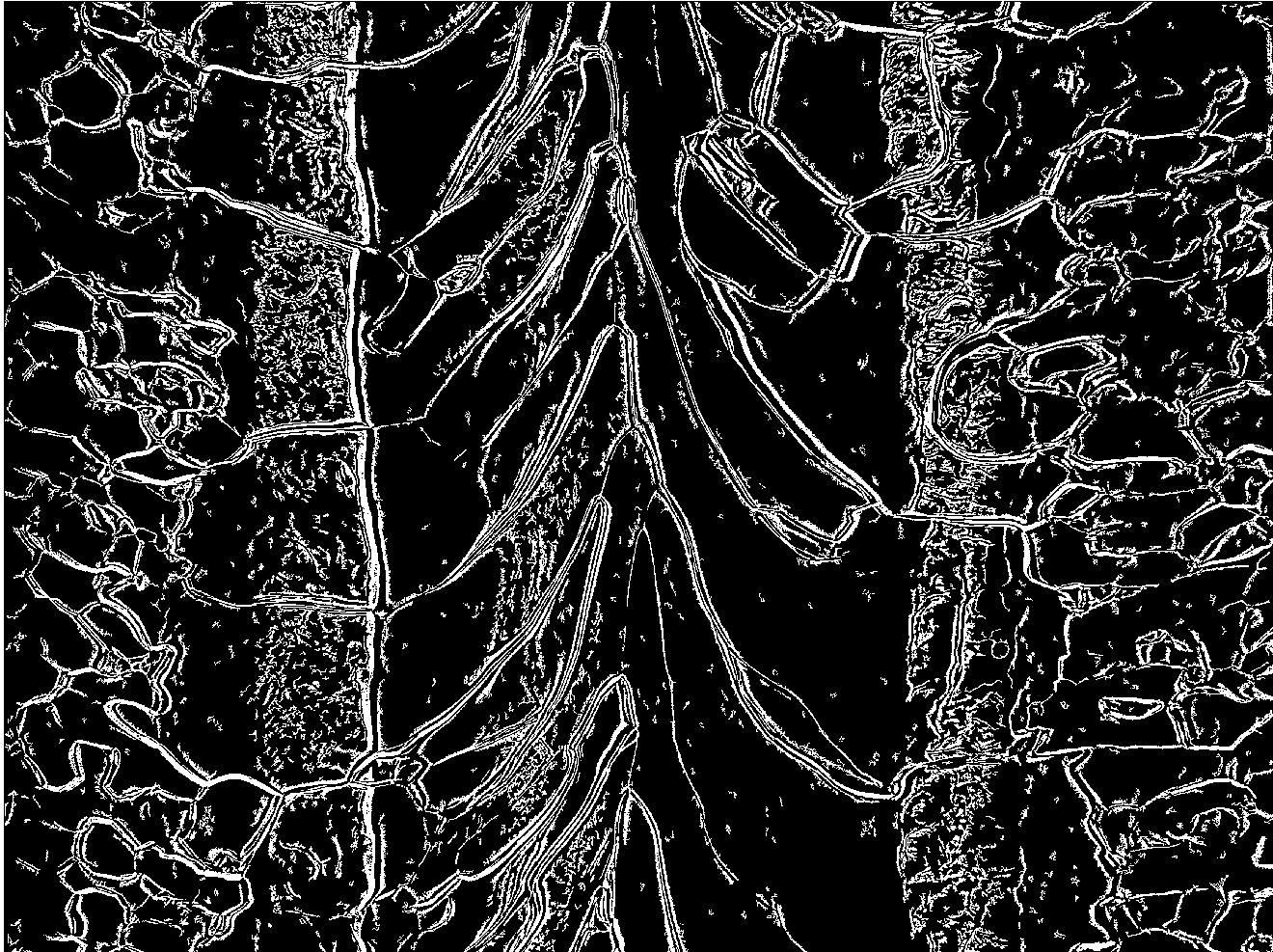}
	\caption{The binary image of the image shown in the previous figure, derived with the image processing algorithm.}
	\label{fig:acbw}
\end{figure}

The output of the image processing algorithm is a binary image (see Figure \ref{fig:Flowchart} for process flow). In Figure \ref{fig:acorigina}, an example of an input image as taken by OBACHT is shown. The output of the algorithm, for this example image, is shown in Figure \ref{fig:acbw}. Both images are transferred to the image analysis code. 

\section{Algorithm Benchmark}
\label{sec:benchmark}
To discuss and interpret results obtained using the presented algorithm, its performance is explored. Several benchmark scenarios were used to gain more information about the performance and limitations of the algorithm. 

\subsection{Resolution}
\label{sec:resolution_usaf}
\begin{figure}[ht]
	\centering
		\includegraphics[width=0.2\textwidth]{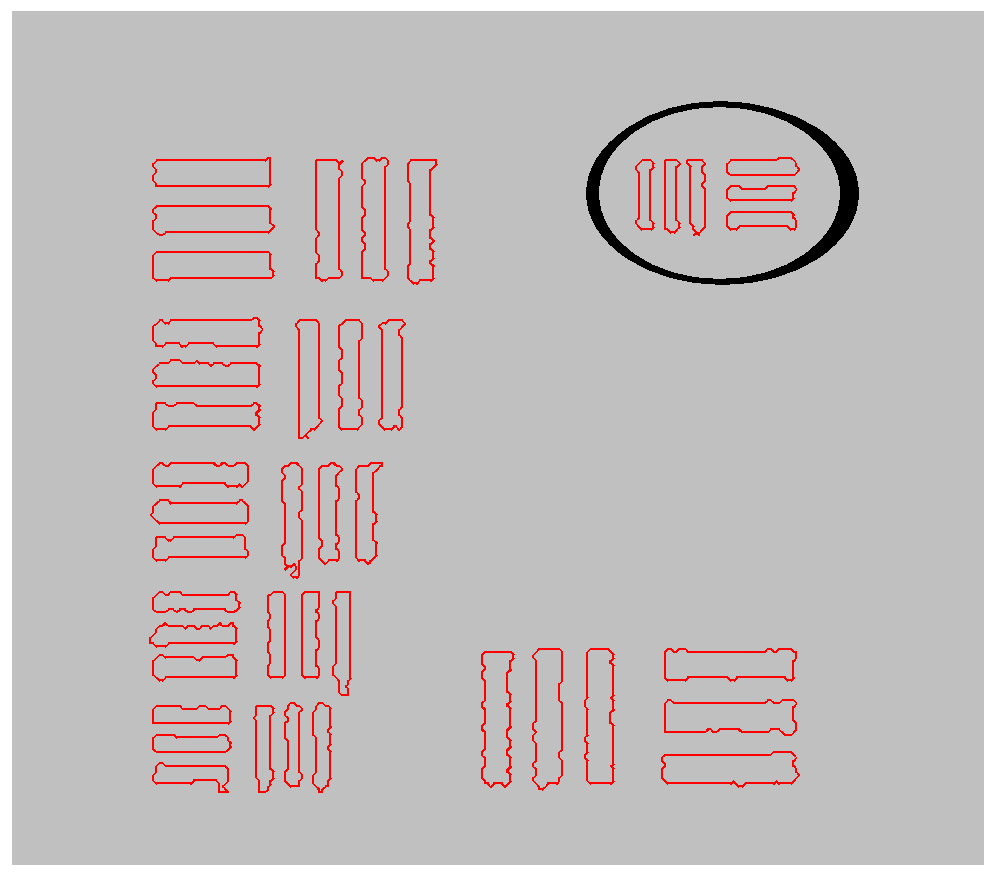}
		\caption{USAF1951 test chart after image processing. The red lines show the edges of the as individual objects detected elements. The smallest separated group elements - encircled - are the elements of group five, elements one. This results in a resolution of 15.63\,$\mathrm{\upmu m}$.}
	\label{fig:Overlay_L3_image}
\end{figure}

To investigate the resolution limit of the image processing algorithm, a set of well known test images was used to identify the smallest distance still resolved.
\newline
The benchmark pattern was the USAF 1951 test chart, the same image as used for the optical resolution in section \ref{section:OSDOF}. The image is processed with the algorithm and the boundary pixels of the detected group elements are shown in Figure \ref{fig:Overlay_L3_image}.
The smallest objects, which are still detected as individual stripes, are part of group five, element one. This results in a algorithm resolution of 15.63\,$\mathrm{\upmu m}$ with good contrast. The algorithm resolution is slightly below the resolution of the optical system (12.4\,$\mathrm{\upmu m}$). This is because of the filtering and smoothing procedure, which tends to connect objects with a distance smaller than four pixels, and due to the area cut to remove shot noise objects.
\newline
Additionally, a contrast dependent resolution check was performed. A series of white lines with a thickness of ten pixels and a distance of 5 to 15\,pixels was placed on 16 different backgrounds. The background intensity was increased in 16\,Bit steps from [0, 240] to test the contrast dependence of the resolution. Lines, spaced by a distance of 10\,pixels or more, which equals 35\,$\mathrm{\upmu m}$ or more on the cavity surface or twice the algorithm resolution, have been resolved independently of the contrast. Lines with a spacing below a distance of 10 pixels had to have an intensity difference to their background of at least 16\,Bit to be resolved. This contrast robustness of the resolution is based on the CLAHE algorithm.

\subsection{Accuracy}
\label{sec:accuracy}
\begin{figure}[ht]
	\centering
		\includegraphics[width=0.6\textwidth]{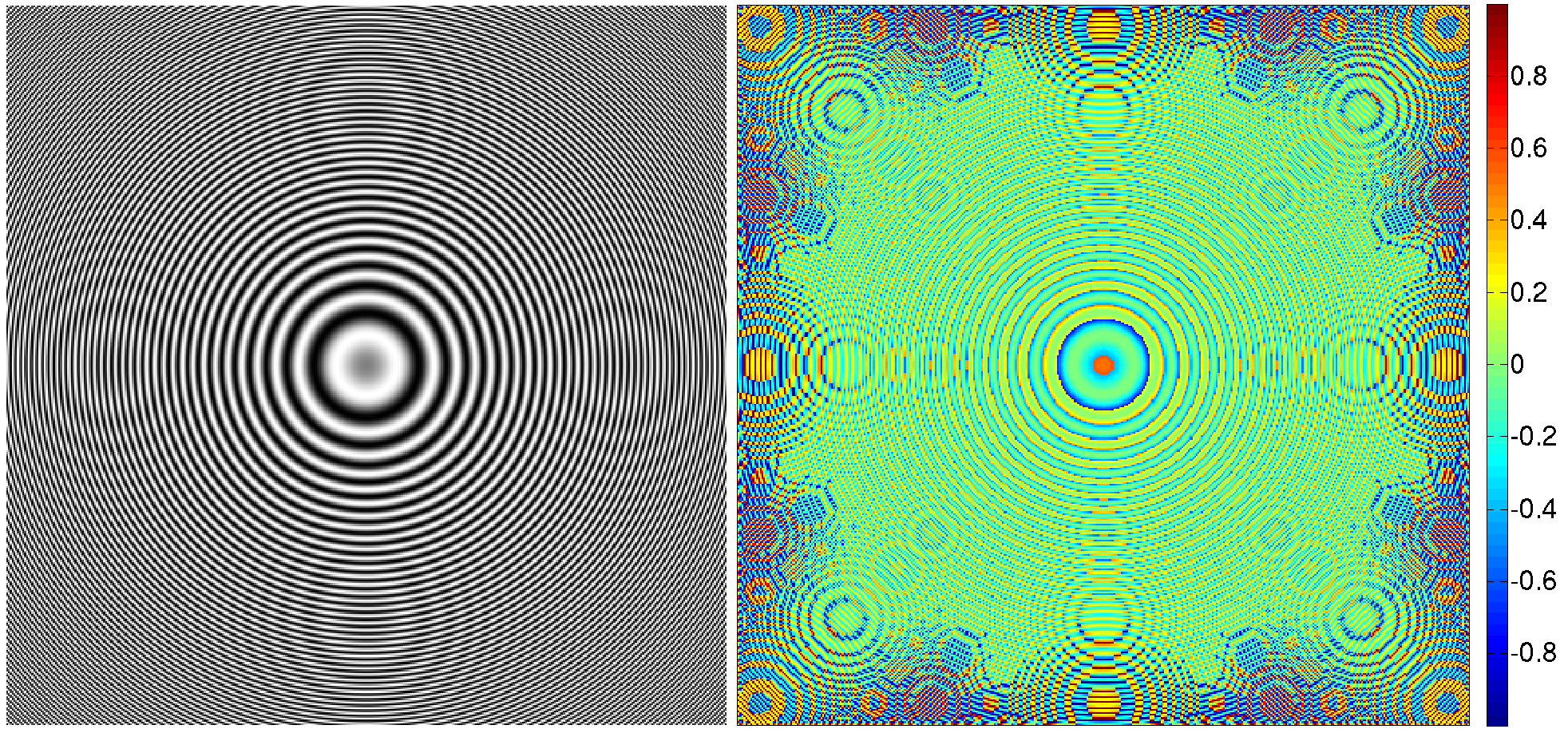}
	\caption{On the left is the J\"ahne test image: A set of concentric rings. Any other pattern seen is due to the aliasing artifact in the low resolution printing. On the right the difference image between the J\"ahne test image and the binary image from the image processing algorithm is shown.}
	\label{fig:Jaehen_Test_Original}
\end{figure}

The image processing algorithm can be interpreted as a classifier, since the binary image classifies each pixel either as a background (no boundary) or a foreground (boundary) pixel. The accuracy off a classifier is defined as 
\begin{equation}
\label{eq:acc}
\mathrm{Accuracy} = \frac{\sum \mathrm{true~positive} + \sum \mathrm{true~negative}}{\sum \mathrm{All~pixels}}
\end{equation}
with true negatives or true positives as pixels which are correctly identified as background or foreground pixels. Also interesting is the positive predictive value (PPV). The definition is
\begin{equation}
\mathrm{PPV} = \frac{\sum \mathrm{true~positive}}{\sum \mathrm{true~positive} + \sum \mathrm{false~positive}}
\end{equation}
and gives a measure of the probability that a pixel identified as boundary pixel is truly a boundary pixel. This value can be used to correct the derived object area from the algorithm. 
\newline
In order to have the ability to decide whether a pixel is rightly or wrongly classified, a test image with known properties is used. Here, the J\"ahne test image $g_1$ \cite{Jaehne2004}, shown in Figure \ref{fig:Jaehen_Test_Original}, is used. It can be calculated via
\begin{equation}
g_0=\frac{\sin(\pi r^2)}{r_{max}}; g_1=\frac{(g_0+1)}{2}.
\end{equation}
With $g_i$ the gray scale test image, $r$ as the radius of the circle and $r_{max}$ the maximum radius, 283\,pixels in this case. This radius was chosen to generate boundary distances similar to a real OBACHT image. The latter equation is used to scale and shift the image to the interval [0,1] Bit, which allows a direct comparison of $g_1$ to the binary image. 

\begin{figure}[htbp]
	\centering
		\includegraphics[width=0.9\textwidth]{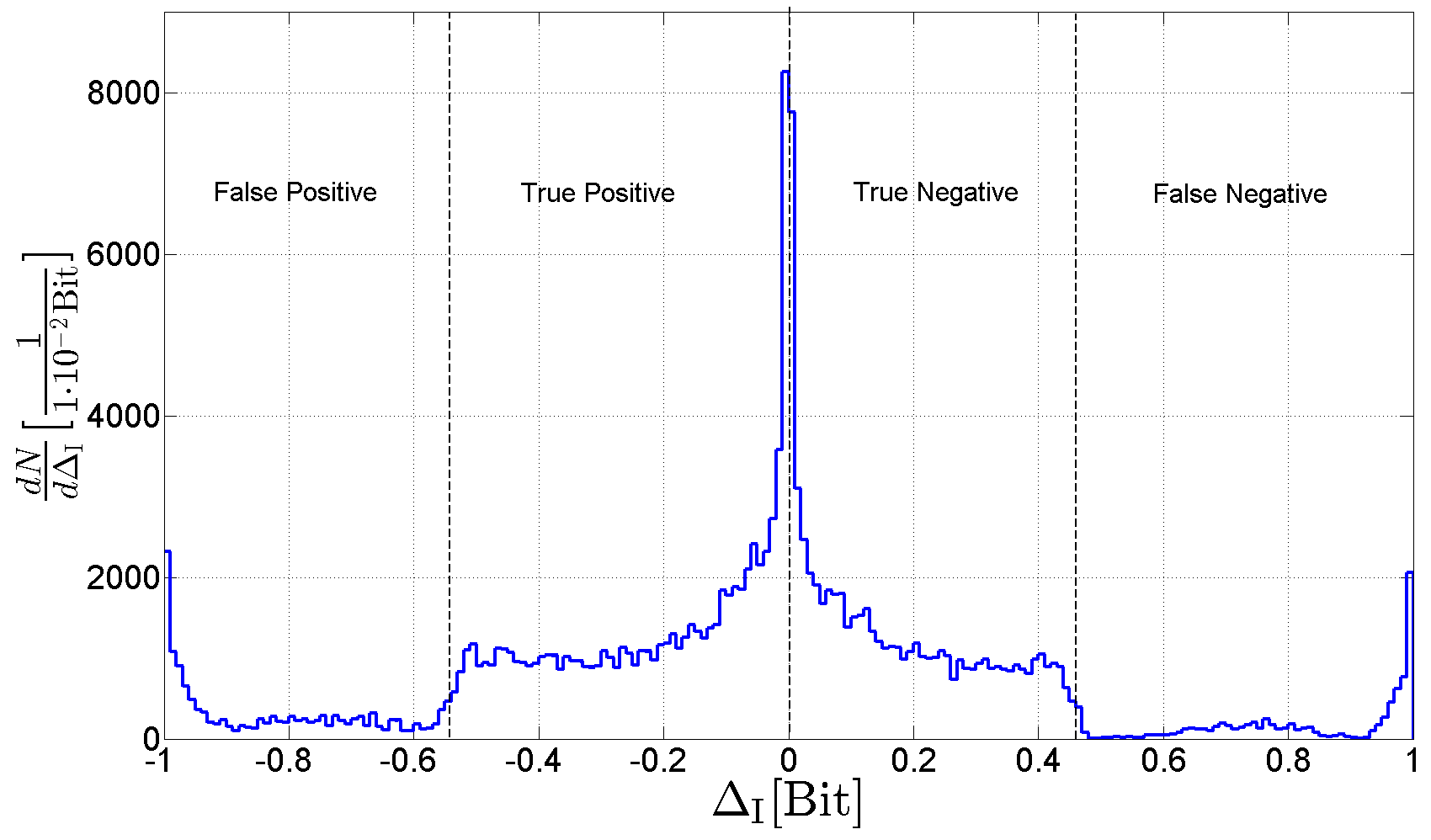}
	\caption{The plot shows the histogram of the pixel value distribution of the difference image. The corresponding classification intervals are depicted by dashed black lines.}
	\label{fig:K_Hist}
\end{figure}

\begin{figure}[htbp]
	\centering
		\includegraphics[width=0.9\textwidth]{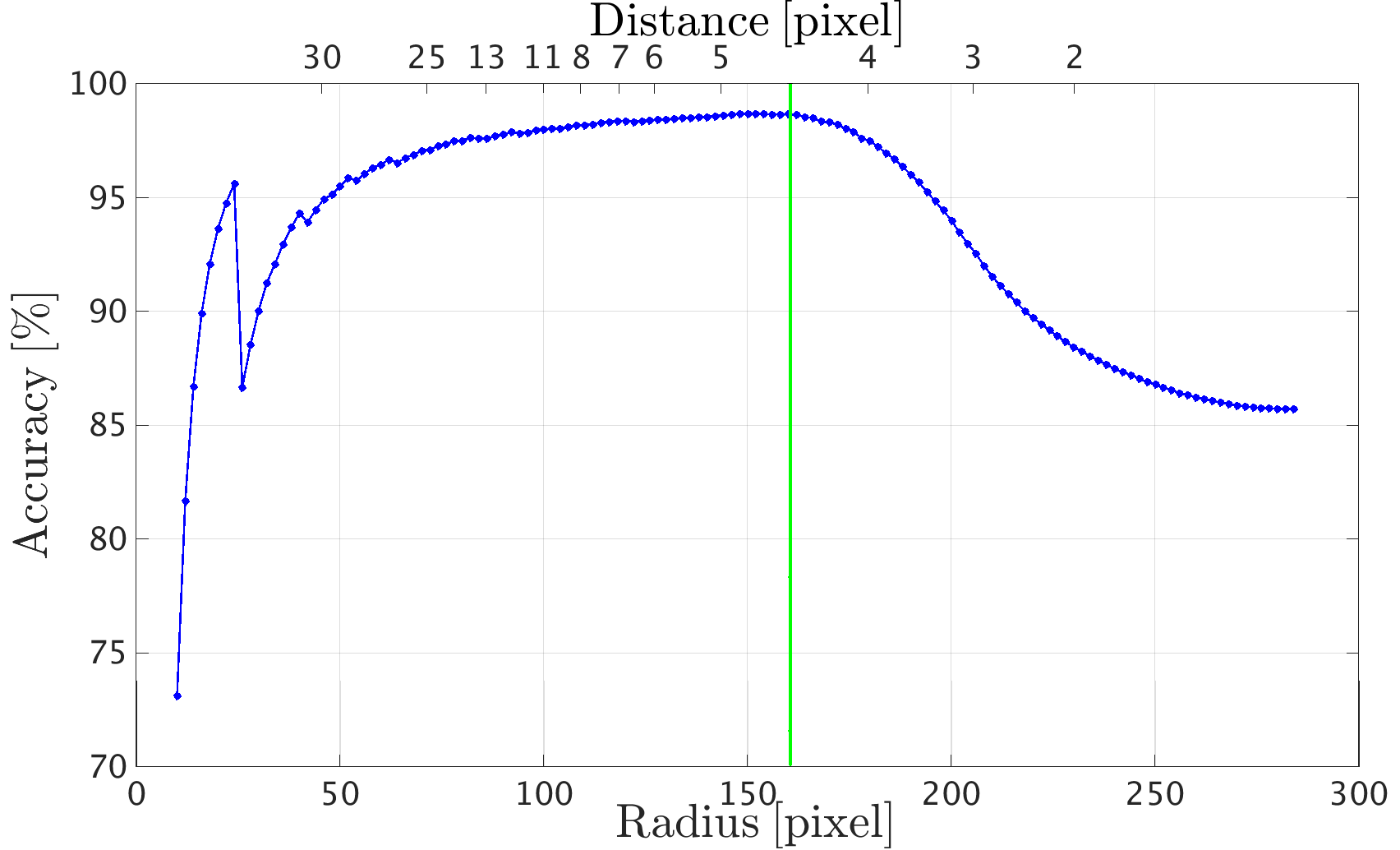}
	\caption{Accuracy as a function of the radius. On the lower x-axis the radius r is shown, while the y-axis shows the accuracy for the corresponding circular region with radius r. The upper x-axis shows the distance between two consecutive boundaries at the corresponding radius. The green line depicts the best value and its corresponding radius.}
	\label{fig:Accuracy_vs_r_Difference_vs_r}
\end{figure}

\begin{figure}[htbp]
	\centering
		\includegraphics[width=0.9\textwidth]{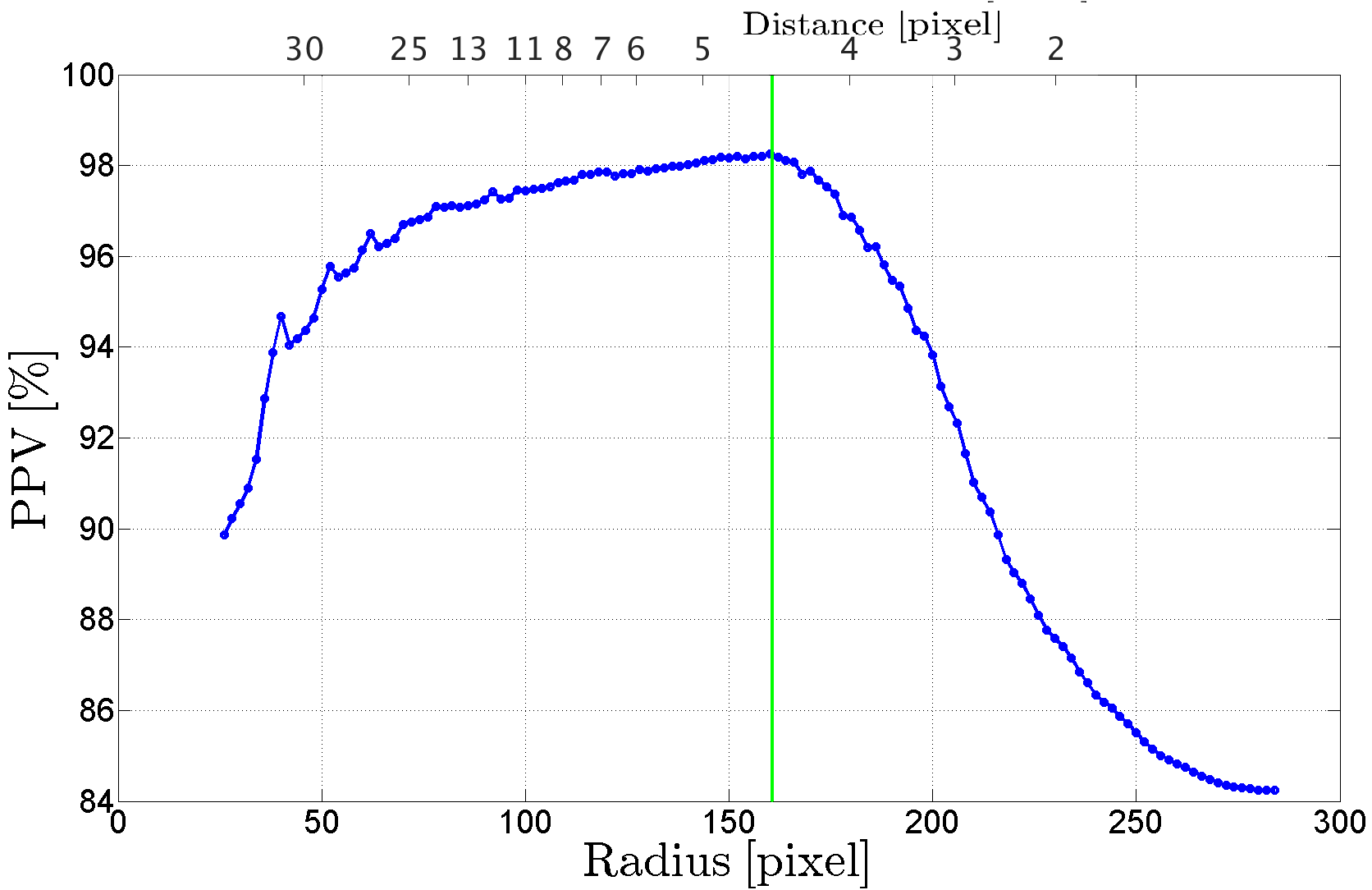}
	\caption{PPV as a function of the radius. On the lower x-axis the radius r is shown, while the y-axis shows the PPV for the corresponding circular region with radius r. The upper x-axis shows the distance between two consecutive boundaries at the corresponding radius. The green line depicts the best value and its corresponding radius.}
	\label{fig:PPV}
\end{figure}

The ring cross sections can be approximated with Gaussian profiles. With this, the $\pm 1 \upsigma$ region of the Gaussian profile is defined as ring width, resulting in the classification that any pixel with a value between [0.46, 1] Bit is part of the ring. 
The image is processed and the resulting binary image is subtracted from the original image. To interpret the difference image, it has to be kept in mind that:
\begin{itemize}
	\item The image is the difference between a continuous Gaussian value ([0,1] Bit) and a discrete step function (0 or 1 Bit) for each pixel.
	\item A true ring pixel has a Gaussian value larger than 0.46 Bit, a true background pixel a value smaller than 0.46 Bit.
\end{itemize}
There are four classifications possible, each identifiable with a distinct intensity interval, see Table \ref{tab:ContTable}.
\begin{table}[ht]
	\centering
	
		\begin{tabular}{lcc}
		\toprule
					& \textbf{Positive} & \textbf{Negative} \\ \cmidrule{2-3}
		\textbf{True}  & F / F = [-0.54,0] Bit & B /B = [0,0.46] Bit				  \\ 
		\textbf{False} & B / F = [-1,-0.54] Bit &F /B = [0.46,1] Bit				\\
	  \bottomrule			
		\end{tabular}
		\vspace{0.2cm}
		\caption{Contingency table with corresponding intensity regions in difference image. F = Foreground, B = Background.}
		\label{tab:ContTable}
\end{table}
With the given contingency table, it is now possible to calculate the mentioned performance characteristics of the algorithm. The difference between the J\"ahne test image and the binary image is also shown in Figure \ref{fig:Jaehen_Test_Original}. The respective histogram of the intensity distribution of the difference image is shown in Figure \ref{fig:K_Hist}.
The peaks observed in the histogram at $\pm 1$ and 0 Bit is due to the circular shape of the test pattern. 
With the given intensity distribution, the relative probabilities of the classification cases can be calculated, see Table \ref{tab:ContTable_relProb}. The accuracy and the PPV of the algorithm can be calculated to be 85\,\% respectively 84\,\%.
\begin{table}[ht]
	\centering
	
		\begin{tabular}{lcc}
		\toprule
					& \textbf{Positive} & \textbf{Negative} \\ \cmidrule{2-3}
		\textbf{True}  & 47\,\% & 38\,\%			  \\ 
		\textbf{False} & 9\,\% &	6\,\%				\\
	  \bottomrule			
		\end{tabular}
		\vspace{0.2cm}
		\caption{Contingency table with relative probability.}
		\label{tab:ContTable_relProb}
\end{table}
But these values show a dependency on the radius. Rings with a large radius have a small distance to the neighboring rings by definition. This can lead to two problems in the processing
\begin{enumerate}
	\item The slope of the Gaussian profile is too steep. This will introduce an aliasing effect due to the sampling of the profile on a discrete grid. This leads to a wrong classification of the pixel, either false positive or false negative depending on the phase between the slope and the grid and the value of the slope. 
	\item The distance between two rings is to small. The filtering and smoothing procedure during image processing merges rings and result in a false positive classification. 
\end{enumerate}
These problems are similar to the pixel noise consequences (2) and (3). In Figure \ref{fig:Accuracy_vs_r_Difference_vs_r}, the accuracy is plotted as a function of the radius. The accuracy clearly decreases above a radius of 160\,pixels. Within this region with r = 160\,pixels, the accuracy is 97\,\%. Rings with a radius larger than 160\,pixels have a distance of four pixels or less and therefore also a high slope. This is the reason for the decrease of accuracy and shows a resolution limitation similar to the result in section \ref{sec:resolution_usaf}, where the USAF test chart shows also a drop in resolution if edges are closer than four pixels. There are two reasons for an accuracy below 100\,\% at a radius below 160\,pixels where both are caused by the very test pattern. The first reason is the region with a radius below 18\,pixels. This region is identified as background with image processing code, but defined as foreground since its intensity values are above 0.46\,Bit and leads to a false positive classification. The second reason, which is also the cause for the drop in the accuracy at 26 pixels, is the region between the radius of 20 to 30 pixels. A sudden increase in pixels classified as false positive is visible. 
\newline
Figure \ref{fig:PPV} shows the PPV of the algorithm as a function of the radius. The PPV within a radius of 160\,pixels is 97\,\%.
The developed image processing algorithm, optimized for OBACHT images, shows a maximal resolution of 15.63\,$\mathrm{\upmu m}$ close to the optical system, which is 12.4\,$\mathrm{\upmu m}$. The accuracy within the resolution limit of the algorithm is 97\,\%, PPV 99\,\%. The actual values for the PPV and the sensitivity for an OBACHT image are reduced by the shot noise. Adding a gaussian noise with a variance as in the OBACHT images and a mean of zero to the J\"ahne test image reduces the accuracy to 75\% and the PPV to 87\%. The values for the image processing code of an OBACHT image will be between these two cases.

\section{Definition of Variables}
\label{sec:def_variables}
After processing the images, the characteristics of the boundaries can be measured. This procedure is part of the image analysis algorithm. The input of this algorithm is the binary image, including the linked map, produced by the image processing algorithm and the original image. Only variables used in this paper are presented.

\subsection{Boundary Area}
\label{sec:obj_area}
First, the area projected on a single pixel has to be calculated. This is quite simple, since the pixel length on sensor is known to be 1.75\,$\mathrm{\upmu m}$ and the magnification M was calculated to be 0.51, see section \ref{section:OSDOF}. This means, that a surface length of 3.5\,$\mathrm{\upmu m}$ is projected on a single pixel. The area seen per pixel, or simple "pixel area", is $3.5 \times 3.5\,\mathrm{\upmu m^2}$. 
\newline
The area of a boundary is simply the number of pixels of this object times the area of a pixel. A more elaborate definition can be found using the image moment $m_{p,q,i}$ \cite{Hu1962}
\begin{equation}
\label{eq:4.3}
m_{p,q,i}=\sum^{M}_{x=1}\sum^{N}_{y=1} x^{p} y^{q} ImF\left(x,y,i\right)
\end{equation}
where $ImF\left(x,y,i\right)$ is the image function, which defines whether a pixel at coordinates $(x,y)$ is part of the boundary with the label $i$ or not, where the value of this function is zero or one respectively and the information is taken from the labeled binary image. M is the number of columns and N the number of rows of the image. The parameters p, q are any positive integers and define the order of the moment as $\left(\mathrm{p} + \mathrm{q}\right)$. The coordinate origin is the lower left corner of the image.
In this notation, the area A of the boundary i is simply the zero-th order moment $m_{0,0,i}$ times the pixel area. Although, this notation seems artificially complicated for this situation, its use will be seen in the context of the other variables described next.
\newline
With the given resolution at OBACHT, the experimentally obtained relative error for the grain boundary area is 3\,\%, similar to \cite{Patil2011}.

\subsection{Boundary Centroid}
\label{sec:dtm}
The centroid position of an object, using the image moment notation, is 
\begin{equation}
\vec{R}= \left( \frac{m_{1,0}}{m_{0,0}}, \frac{m_{0,1}}{m_{0,0}} \right).
\end{equation} 
with the components 
\begin{equation}
\bar{x} = \frac{m_{1,0}}{m_{0,0}} \,\ , \bar{y} = \frac{m_{0,1}}{m_{0,0}}.
\end{equation}

The component $\bar{x}$ is used to derive the distance of the object from the image axis and to classify an object as part of the welding seam, the heat affected zone or the bulk niobium. The $\bar{y}$ component represents the azimuthal position of an object in the cavity coordinate system. The uncertainty of the centroid under pixel noise is quite stable. With the formalism developed in \cite{Klette1999}, an upper limit of the systematic uncertainty of 2 pixels in each direction of the centroid is obtained.

\subsection{Boundary Orientation}
\label{sec:orient_var}
\begin{figure}[htbp]
	\centering
		\includegraphics[width=0.4\textwidth]{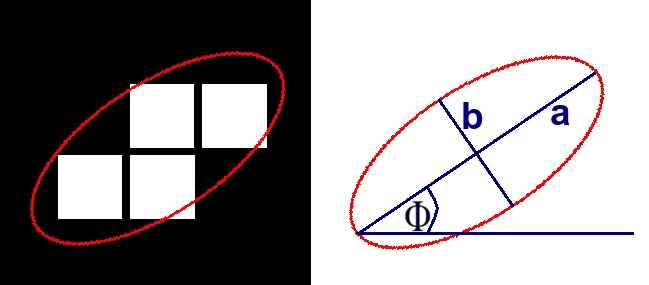}
	\caption{On the left side, a four pixel object and the ellipse with the same second central moment is shown. On the right side, the ellipse with its major axis a and minor axis b and the horizontal z-axis for the angle assignment $\Phi$ is shown.}
	\label{fig:ellipse}
\end{figure}

Objects are not 'a priori' symmetric. Hence, it is nontrivial to define properties like diameter or major axis length, eccentricity and orientation of an object. One method to do this, is to define an ellipse which has the same second central moment as the pixel distribution of the pixels \cite{Hu1962}, as the object consists of, cf. Figure \ref{fig:ellipse}. 
\newline
To calculate the variables mentioned above, the central moments $\upmu_{p,q}$ of the object need to be calculated
\begin{equation}
\upmu_{p,q,i}=\sum^{M}_{x=1}\sum^{N}_{y=1} \left(x-\bar{x} \right)^p \left(y-\bar{y} \right)^q ImF\left(x,y,i\right)
\end{equation}
with $\bar{x}$ and $\bar{y}$ as the x- and y-component of the centroid and $ImF\left(x,y,i\right)$ as the image function. These central moments can be calculated directly from the already introduced image moments.
The orientation of an object with respect to the x-axis is calculated via
\begin{equation}
\Theta = \frac{1}{2} \arctan \left( \frac{2\upmu'_{1,1}}{\upmu'_{2,0}-\upmu'_{0,2}} \right)
\end{equation}
with 
\begin{equation}
\upmu'_{p,q}=\upmu_{p,q}/\upmu_{0,0}. 
\end{equation}
With the formalism in \cite{Klette1999,Liao1993}, an upper limit of the systematic uncertainty of $5^\mathrm{o}$ is derived. A circle, with equal moments $\upmu'_{2,0}$ and $\upmu'_{0,2}$, would lead to an ill-defined angle since the denominator becomes zero. For this case the orientation is defined as $0^\mathrm{o}$.

\subsection{Boundary Roughness}
The surface roughness should be reflected in the intensity variations observed in the image. Hence we assume that the intensity gradient can be used as a measure for the surface roughness, which is actually the underlying principle of the boundary detection from the beginning.
\label{sec:rdq}
\paragraph{Intensity Gradient} \hfill \\ 
The intensity gradient $\Delta$ for the gray scale image matrix is calculated for each dimension, $x$ and $y$ using finite differences:
\begin{eqnarray}
\Delta_{x}\left(i,j \right)  = I\left(i+1,j \right) - I\left(i-1,j \right) 
\label{eq:int_diff}
\end{eqnarray}
and in the same manner for $y$ with $\mathrm{I}$ being the gray scale image were $i$ and $j$ are particular values of the pixel coordinates in $x$ and $y$ direction. The finite difference is divided by the step size, and the unit for this obtained variable is $\frac{\mathrm{Bit}}{\mathrm{\upmu m}}$. The intensity is represented in 256 Bit while the step size from pixel to pixel is 3.5\,$\mathrm{\upmu m}$. 
This operation generates two new image matrices, $g_x$ and $g_y$. These images have the same dimension as $\mathrm{I}$ and contain the gradient values in $x$ and $y$ direction. To obtain the scalar magnitude G of the image gradient, the dot product of the vector ($g_x\left(i,j \right),g_y\left(i,j \right)$) is calculated.

\paragraph{Surface slope} \hfill \\ 
Starting with the intensity gradient, a quantity called $\mathrm{R_{dq}}$ - the boundary or surface roughness in this work - is introduced. It is based on ISO 25178 for surface texture \cite{iso25178} and represents the RMS of the slope of the object within the sampling length. 
$\mathrm{R_{dq}}$ is the root mean square of the intensity gradient within the sampled area of the boundary. A steeper slope of a boundary or surface would imply a larger intensity gradient and hence a larger $\mathrm{R_{dq}}$. The $\mathrm{R_{dq}}$ can be calculated for each identified bounded object from:
\begin{equation}  
\mathrm{R_{dq}}=\sqrt{\frac{1}{n}\sum_{x=1}^M\sum_{y=1}^N G^2\left(x,y \right)ImF\left(x,y,i \right)}
\end{equation}
with $ImF\left(x,y,i \right)$ the image function and $n$ the number of pixels of the boundary.

\paragraph{Uncertainty} \hfill \\
A statistical noise arises from the $SNR$ of the image sensor in the camera, which is 32\,dB. The $SNR$ of a digital image can be calculated with \cite{young1998}
\begin{equation}
SNR = 20 \mathrm{log}_{10} \left(\frac{I_{max}-I_{min}}{\upsigma_{bg}}\right)
\end{equation}
with $\upsigma_{bg}$ as the standard deviation of the noise with a zero mean and $I_{max}$ and $I_{min}$ the highest and lowest intensity value within the image. $I_{max}$ is on the order of 0.9 Bit and $I_{min}$ on the order of 0.1. Hence, $\upsigma_{bg}$ is 0.02 Bit. With given definition of $\mathrm{R_{dq}}$, its uncertainty can be calculated and yields to $\upsigma_{R_{dq}} = \frac{0.011}{\sqrt{n}}\,\frac{\mathrm{Bit}}{\upmu \mathrm{m}}$  where $n$ is the number of pixels enclosed in a given object. 
\newline
A systematic uncertainty due to image focus blurring was found to be  $\frac{\delta \mathrm{R_{dq}}}{\mathrm{R_{dq}}}= 3\,\%$. For more details on the image processing algorithm and explicit definitions and discussion of the obtained variables see \cite{Wenskat2015}.

\section{Variable Benchmarks}
Two variables introduced to asses the SRF performance from measurements of surface images, namely boundary area and boundary roughness, are now discussed concerning their effectiveness and plausibility.
\subsection{Boundary Area}
\label{sec:benchmark_area}
\begin{figure}[htbp]
	\centering
		\includegraphics[width=0.3\textwidth]{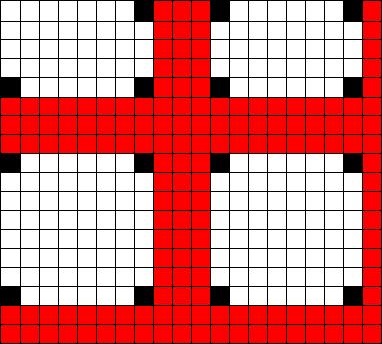}
	\caption{Exemplary test pattern with squares of a side length of 8 with false positive pixels. The red pixels are true boundary pixels. Black pixels are false positive pixels of the boundary because of a higher intensity gradient than the average background pixels.}
	\label{fig:Testpattern}
\end{figure}

As defined in section \ref{sec:obj_area}, the area of a boundary is simply the number of pixels of this object times the area of a pixel. Since the artificial test patterns used here are not images taken with OBACHT, the area of a pixel would be an artificial number and is neglected. Hence, the area of a boundary is just a number of pixels for this consideration.
\paragraph{Test Pattern} \hfill \\ 
The used test patterns are two different pixel grids. Given the geometry of the test pattern - total area of 300 $\times$ 300\,pixels and boundaries with a thickness of 3 pixels and a total number of lines of n=11 and 5 with a spacing of 22 and 47\,pixels respectively - the area of each object can be calculated.
The area of a square is $22^2 = 484$\,pixels respectively $47^2 = 2209$\,pixels. The average area of a square, as found by the algorithms, is 483 pixels and 2204 pixels. The slightly smaller numbers found by the algorithm can be explained with the design of the image processing algorithm. The algorithm was designed to detect the boundaries in an image, which are defined as pixels with an intensity gradient above a local threshold. Some pixels close to an intersection of two boundaries can be falsely identified as part of the boundary, as shown in Figure \ref{fig:Testpattern}.
\newline
The formula for the boundary area A is
\begin{equation}
A = 2\cdot(n \cdot 3 \cdot 300)-(n^2 \cdot 9)+(300 \cdot 2) - (2\cdot n \cdot 3) 
\end{equation}   
where $n$ represents the number of lines in one direction. The first term describes the total area covered by the lines in both directions. The second term corrects the area for the double-counted area of the intersections of the grid lines. The third term adds the 2 pixels thick line at the right and bottom edge. The fourth term corrects the area for the double counted area of the intersection between the grid lines and the lines at the two edges. The area of the boundary can be calculated to be 9345\,pixels respectively 19245\,pixels. 
\newline
The area of the boundary, as found by the algorithms, is 9488\,pixels and 19820\,pixels. The values are 1.5\,\% and 2.9\,\% respectively above the true area or 143\,pixels and 575\,pixels in absolute numbers. Since boundary pixels are pixel above a certain intensity gradient, this observed difference can be calculated for the test pattern. Assuming that the false positive pixels are the outer corner pixels at the intersections of lines or with an edge, c.f. Figure \ref{fig:Testpattern}.
The total number of false positive pixels can be calculated via
\begin{equation}
A = (n^2 \cdot 4) + (2 \cdot n \cdot 2) + (2 \cdot n \cdot 2) + 3
\end{equation}   
where n represents the number of lines in one direction. The first term are the false positive pixels at the intersections of two grid lines. The second and third term represents the false positive pixels created at the intersection of the grid lines with the edges of the image. The fourth, constant, term is the contribution of the 2 pixels thick lines at the right and bottom image edge. For the 25\,and 50\,pixels grid, the number of false positive pixels are 143\,and 575\,pixels, which equals the difference of the area found by the algorithm and the true area.
\newline
Summing up, the boundary area as derived by the image processing algorithm behaves as expected. The relative error is in the same order of magnitude as found with the image processing accuracy benchmark.
 
\begin{figure}[htbp]
	\centering
		\includegraphics[width=0.75\textwidth]{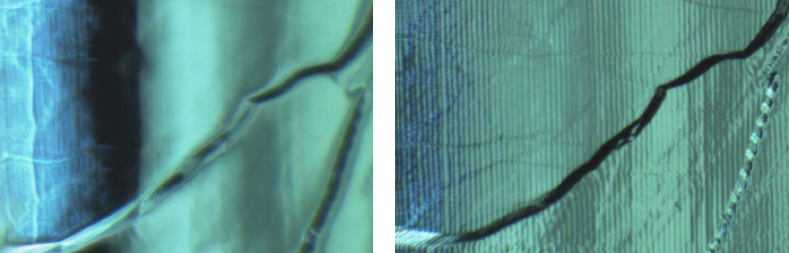}
	\caption{OBACHT image of the same boundary after EP (left) and BCP (right) \cite{Sebastian}.}
	\label{fig:BoundaryExamples}
\end{figure}

\begin{figure}[htbp]
	\centering
		\includegraphics[width=0.75\textwidth]{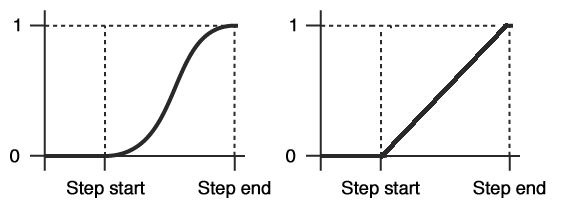}
	\caption{A sketch of a grain boundary and how its shape would change under different treatments: left EP - right BCP. More pixel on the right exceed the intensity gradient threshold than pixel on the left, hence the boundary area is larger.}
	\label{fig:Scheme_Boundary}
\end{figure}

\paragraph{Grain Boundary} \hfill \\ 
The area of the same grain boundary with different visibility due to different surface treatments \cite{Sebastian,Singer2013}, as shown in Figure \ref{fig:BoundaryExamples}, is obtained. The single grain boundary under investigation is running from the lower left to the upper right corner.
After the image processing, the area of the boundary was derived and is shown in Table \ref{tab:AreaExamples}.
\begin{table}[ht]
	\centering
\begin{tabular}{lcc}
	\toprule
	& \textbf{EP} & \textbf{BCP} \\ \cmidrule{2-3}
 	\textbf{Area $\left[\mathrm{mm}^2 \right]$} & 0.05 & 0.10 \\
 	\bottomrule
 	\end{tabular}
	
 	\caption{Area of the same grain boundary after different treatments as found by the algorithm.}
 	\label{tab:AreaExamples}
\end{table}
The grain boundary area between the two surface treatments differ by almost a factor two. The different surface treatments result in different grain boundary slopes, where the electro-polishing (EP) creates a smoother grain boundary than buffered chemical polishing (BCP), see Figure \ref{fig:Scheme_Boundary}.
Hence, a correlation between the boundary roughness and the visible boundary area exist. However they are not exchangeable variables, since the visible boundary area is not only a function of the boundary slope as well as its width and length.

\subsection{Boundary Roughness}
The variable $\mathrm{R_{dq}}$ defined in section \ref{sec:rdq} represents one aspect of the general property of surface roughness, namely the local slope of a grain boundary step. This definition of surface roughness is often used to describe repetitive roughness profiles, which is the case for the grain boundary structure. 
\paragraph{Empirical Interpretation} \hfill \\ 
The derived boundary roughness parameter $\mathrm{R_{dq}}$ is only an indirect measurement of the surface slope. Hence, only a relative statement on the surface texture can be given with this parameter. As an example, the boundary which was already shown in Figure \ref{fig:BoundaryExamples} is used and $\mathrm{R_{dq}}$ has been calculated, and results are given in Table \ref{tab:RdqExamples}.
\begin{table}[ht]
	\centering
\begin{tabular}{lcc}
	\toprule
	& \textbf{EP} & \textbf{BCP} \\ \cmidrule{2-3}
 	\textbf{$\mathrm{R_{dq}}$ $\left[ 10^{-3} \frac{\mathrm{Bit}}{\upmu \mathrm{m}}\right]$} & 7.3 & 11.6 \\
 	\bottomrule
 	\end{tabular}
	
 	\caption{Roughness parameter of the same grain boundary after different treatments as found by the algorithm.}
 	\label{tab:RdqExamples}
\end{table}
The boundary slope for the EP treated grain step was found to be smaller than the slope for the BCP treated grain step. This is in agreement with the expected surface topography after these treatments and with the discussion of the observed boundary area.

\begin{figure}[htbp]
	\centering
		\includegraphics[width=0.75\textwidth]{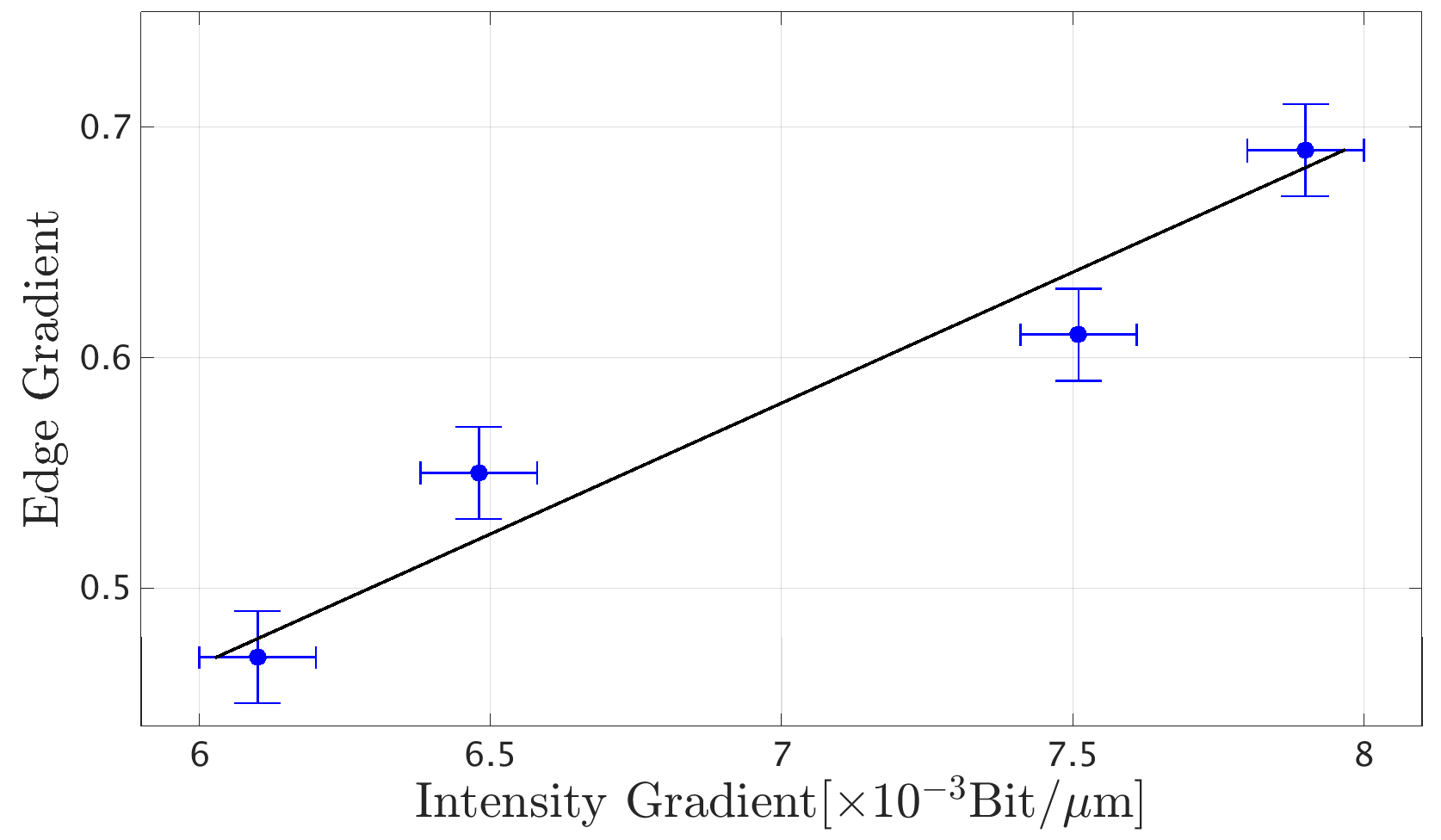}
	\caption{The x-axis shows the intensity gradient of a boundary derived from the image and the y-axis the geometrical gradient measured with the profilometer. The linear fit is also shown.}
	\label{fig:Edgecalibration}
\end{figure}

\begin{figure}[ht]
	\centering
	\includegraphics[width=0.5\textwidth]{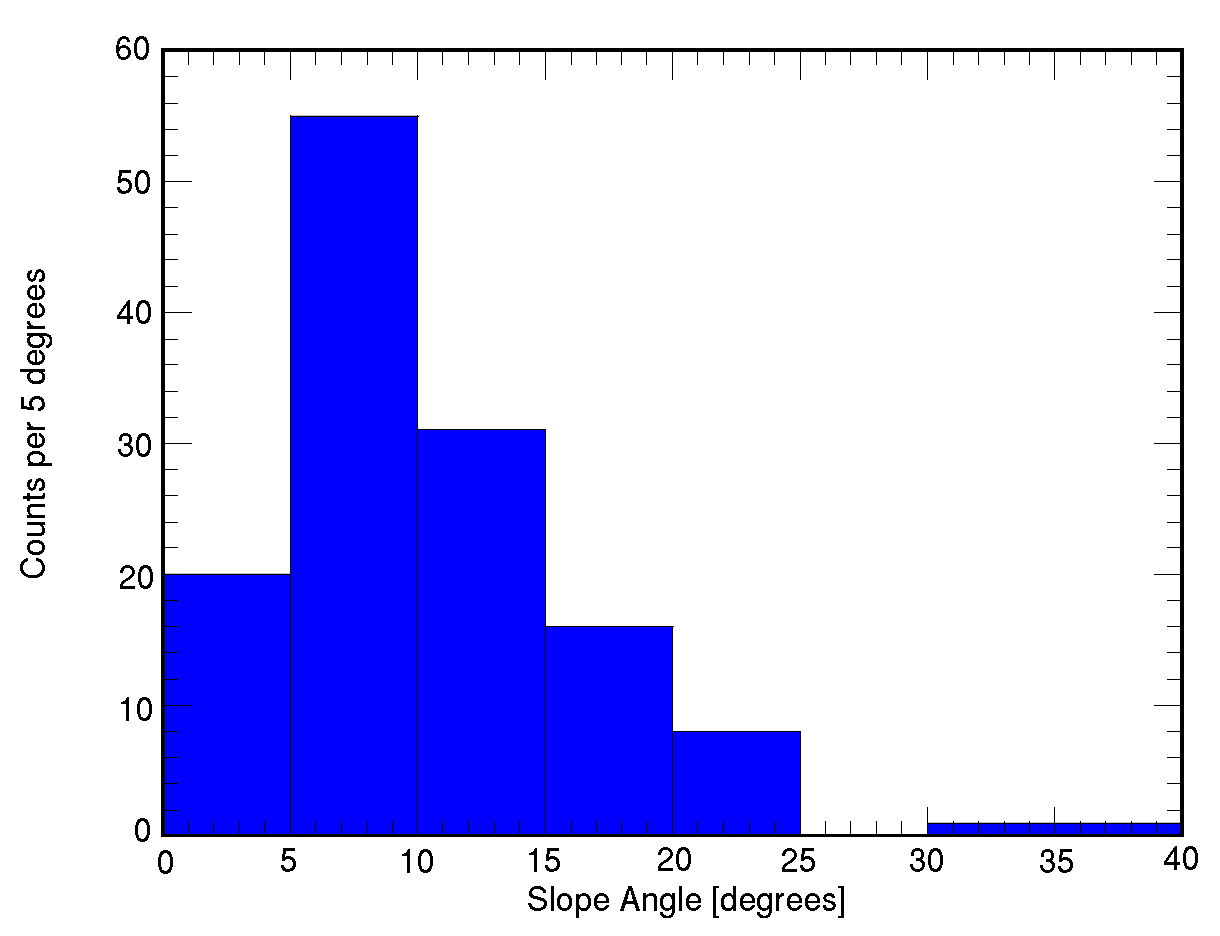}
	\caption{Slope angle histogram obtained from a height profile along a line \cite{Knobloch1999a}.}
	\label{fig:knoblochhist}
\end{figure}

\begin{figure}[htbp]
	\centering
	\includegraphics[width=0.5\textwidth]{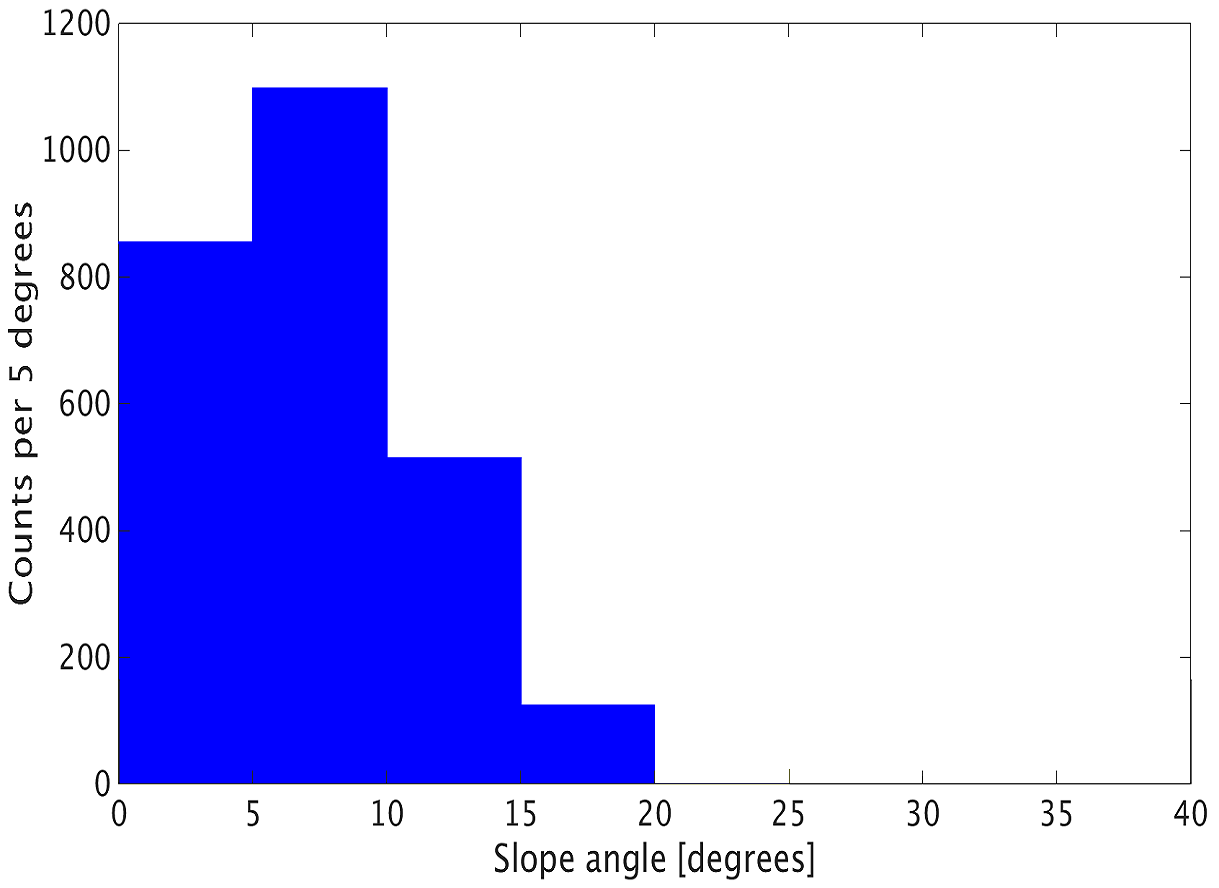}
	\caption{Slope angle histogram obtained from the intensity gradient distribution.}
	\label{fig:eigeneshist}
\end{figure}

\paragraph{Comparison with Profilometric Results} \hfill \\ 
\label{sec:rdq_vs_profilometric}
If the assumption that $\mathrm{R_{dq}}$ is proportional to the surface roughness is true, a correlation between profilometric and optical roughness measurements should be found. To deduce a conversion factor, a replica of a cavity surface was made and scanned with a laser profilometer (UBM Microfocus Expert). The step size of the profilometric scan was 3.5\,$\mathrm{\upmu m}$ in x- and 12.5\,$\mathrm{\upmu m}$ in y-direction. The lateral resolution is 0.06\,$\mathrm{\upmu m}$. A set of 934 $\times$ 334 data points were taken and an area of 3269 $\times$ 4175\,$\mathrm{\upmu m}^2$ was covered. Four boundaries were identified in the area scanned with the laser profilometer. The gradient was measured for each pixel and averaged for the whole boundary. The same boundaries were located in the optical image and the intensity gradient was measured with the image processing and analysis algorithm. In Figure \ref{fig:Edgecalibration} the measured edge gradient and the algorithm deduced intensity gradient are compared, together with a linear fit for calibration.
A conversion function $f$ was derived:
\begin{equation}
f(x) = (113.58 \pm 9.95) \frac{\mathrm{ \upmu m}}{\mathrm{Bit}} \times x - (0.06 \pm 0.07) 
\end{equation}
With this fit, the geometrical slope of the boundaries in an image can be calculated. In Figure \ref{fig:knoblochhist} and \ref{fig:eigeneshist}, a comparison is shown. The upper histogram is derived from a height profile along a line of a niobium sample, after the removal of 120\,$\mathrm{\upmu m}$ with BCP \cite{Knobloch1999a}. The lower histogram depicts a single image of a cavity with the same treatment as the sample in the paper taken with OBACHT and analyzed with the algorithm. 

Although the peak of both distributions are in the same bin and the distributions look similar, the relative counts are different. The different magnitudes of the absolute counts are due to the different sample sizes. The rather hard cut-off at $20^\mathrm{o}$ on the lower histogram can be explained with the resolution limit of OBACHT with given illumination set up, see section \ref{sec:licht}.
\newline
The reason for the different relative counts is the different sensitivity of the two approaches for small angles with respect to each other. The profilometric approach used in \cite{Knobloch1999a}, can only measure steps larger than a few micrometers. This means, a general underestimation of the amount of small angles is inherent in the method. On the other hand, an overestimation of the amount of small angles in the here discussed optical approach is inherent due to shot noise. Shot noise creates intensity fluctuations, even in homogeneous intensity regions and therefore an increase in the amount of small angles. 
\newline
Still, the most likely value and the maximum value are in good agreement. This is quite remarkable since the conversion factor was measured with an independent sample and $\mathrm{R_{dq}}$ is deduced only with optical methods. With given results, the assumption that $\mathrm{R_{dq}}$ is a variable which is proportional to the cavity surface roughness is justified and that $\mathrm{R_{dq}}$ should be sensitive to the difference in the surface roughness due to different surface treatments. 

\section{Automated Report}
The final result of the image processing and analysis algorithm is a mat-file, which is on average 2\,MB per image and it takes on average of 100\,s in CPU time per image. To minimize the real time delay, each incoming image is processed in parallel, yielding 7 images per minute with our server architecture, and all equator images are analyzed after 100 minutes. After all images are processed, the mat-files are combined per equator and a report is generated automatically. This report is then available as PDF and also uploaded as HTML-file for online access. The report contains information such as object density-, grain boundary area-, roughness-, and grain boundary orientation distribution per equator as well as the angular distributions for each equator. In addition, images which deviate - in a not here defined metric - from the average are also shown in the report. Hence, a first assessment of the cavity is available immediately after or even during the optical inspection.

\section{Application}
For this research, more than 100 cavities underwent subsequent surface treatments, cold RF tests and optical inspections within the ILC-HiGrade research program and the XFEL cavity production \cite{CORDIS,Singer2016}. Only an overview of the results of this research is given. For an extensive discussion, see \cite{Wenskat2015,Wenskat2015a,Yegor2017,NavitskiSRF2015}.
\subsection{Vendor Identification}
\begin{figure}[!htb]
	\centering
\includegraphics[width=0.9\textwidth]{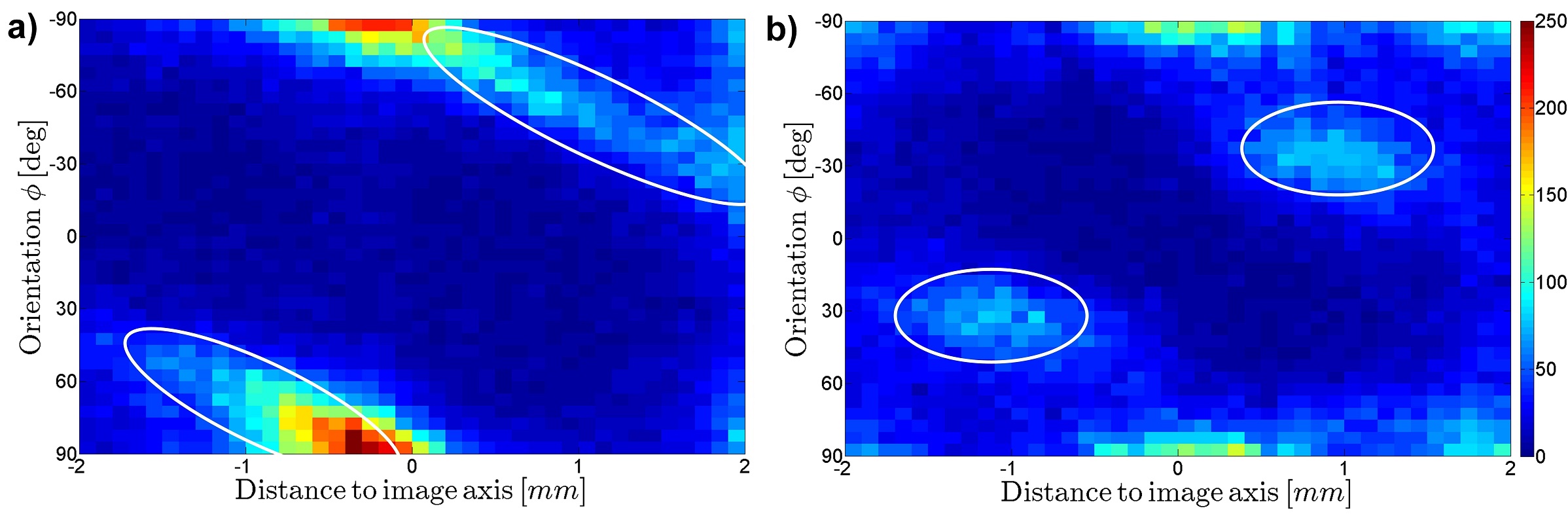}
	\caption{The x-axis shows the boundary centroid position with respect to the image mirror axis, which is the welding seam ridge. The y-axis shows the boundary orientation $\phi$ in degrees w.r.t. an axis perpendicular to the welding seam. Only the welding seam region is shown. The left plot (a) represents boundaries in the welding seam region of a RI cavity, the right plot (b) of an EZ cavity. The white ellipses encircles the welding seam boundaries. The color depicts the counts per bin.}
	\label{Orienthist_vendors}
\end{figure}

As described in \cite{Singer2016, Aderhold2010}, the two cavity vendors, RI Research Instruments GmbH (RI) and Ettore Zanon S.p.A. (EZ), were qualified to produce cavities after two distinct procedures. Most notably is the electron beam welding (EBW) procedure as well as the final surface chemistry step since both have a significant influence on the final surface and therefore onto the RF performance. Since only the standard fabrication procedure should be compared, cavities which underwent any repair were not used in this comparison, which reduces the usable data set since most of the inspected cavities were issued an inspection due to flaws in the fabrication. This data set than yields to a total number of nine RI and eleven EZ cavities which were inspected with OBACHT.  
\paragraph{Electron Beam Welding Procedure}
The solidification dynamics of the weld puddle is under influence of parameters such as temperature gradient, crystalline growth rate and chemical composition. Therefore, the granular microstructure which develops in the weld metal varies and depends on the weld movement pattern, beam travel speed and beam power.
The histograms in Figure \ref{Orienthist_vendors} show the grain orientation in the welding seam region as obtained with the image processing algorithm for each vendor. 
As it can be seen, the boundaries in the welding seam region of EZ have angles of $ \pm 30^\mathrm{o}$. The angles of the boundaries in the welding seam region of a RI cavity show a complete different distribution. At the edge of the welding seam, the boundaries have an angle of $ \pm 30^\mathrm{o}$, similar to EZ, while boundaries towards the center of the welding seam change their orientation. This observed vendor specific grain boundary orientation can be explained by the vendor specific EBW parameters. The travel speed influences the overall bead shape as the shape changes from elliptical to tear drop shaped as the welding speed increases, see \cite{EBW1997, Wenskat2015a}. The grains will grow in the direction of the thermal gradient, starting at the base metal and into the liquified niobium. Comparing this with the observed pattern shows that RI has a beam travel speed slower than EZ, while the exact values are not public. But given the data and experience in \cite{Udomphol2007,EBWBuch}, it can be estimated that the welding speed for EZ has to be on the order of $16\,\pm\,1\,\frac{\mathrm{mm}}{\mathrm{s}}$ while for RI it has to be on the order of $5\,\pm\,1\,\frac{\mathrm{mm}}{\mathrm{s}}$. Those estimated values are in good agreement with available data on welding speeds for different cavities and their grain boundary orientations \cite{Singer2017,Geng1999,Schmidt10}.

\begin{figure}[!htb]
	\centering
		\includegraphics[width=0.9\textwidth]{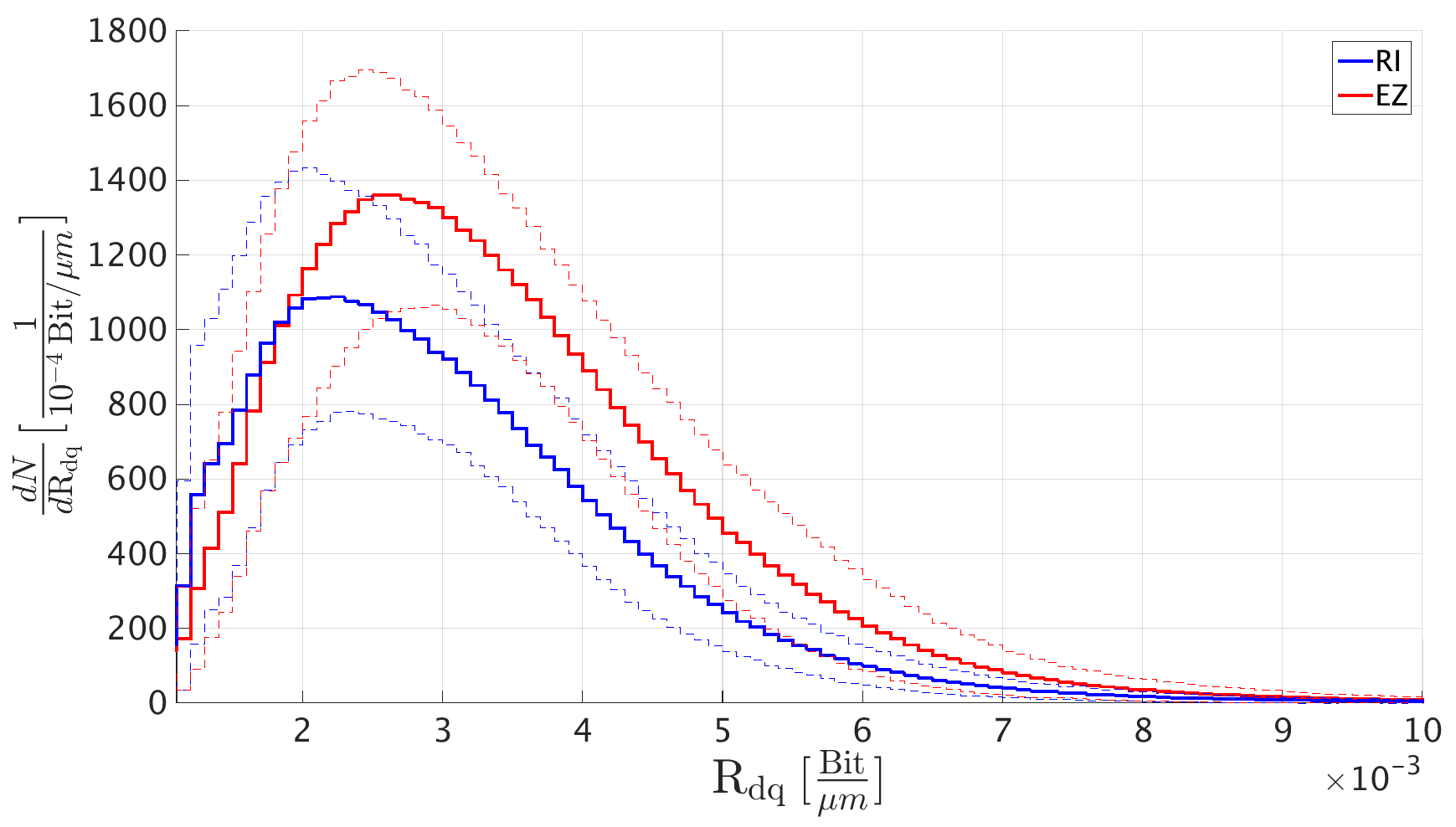}
	\caption{The x-axis shows the grain boundary gradient $\mathrm{R_{dq}}$ for boundaries within the welding seam region and the y-axis shows the counts per bin. The plots show the average $\mathrm{R_{dq}}$ distribution for all cells with a one $\upsigma$ interval. The red distribution is for EZ (N=99) and the blue distribution is for RI (N=81).}
	\label{Vendor_Rdq_Distribution}
\end{figure}

\paragraph{Surface Chemistry}
Cavities from both vendors underwent a bulk EP procedure. In addition, the cavities produced by RI underwent a final EP procedure of 40\,$\upmu \mathrm{m}$ while the EZ cavities underwent a FLASH BCP of 10\,$\upmu \mathrm{m}$. The difference between these surfaces, as parametrized by $\mathrm{R_{dq}}$, is shown in Figure \ref{Vendor_Rdq_Distribution}.
For a quantification, an exponential modified Gaussian distribution (EMG) is fitted to the observed distribution. 

\begin{table}[ht]
	\centering

		\begin{tabular}{l>{\centering\arraybackslash} p{1.8cm}>{\centering\arraybackslash} p{1.8cm}}
		\toprule
		\multicolumn{1}{c}{ }           &    \multicolumn{2}{c}{\textbf{\normalsize{$\overline{\mathrm{R_{dq}}}$}} }   \\
		$\left[ 10^{-3} \frac{\mathrm{Bit}}{\upmu \mathrm{m}}\right]$  & \textbf{WS} & \textbf{HAZ}    \\ 
		\cmidrule{2-3}
		\textbf{RI}  & $2.8 \pm 0.1  $  & $3.1 \pm 0.1 $  \\ 
		\textbf{EZ}  & $3.4 \pm 0.1  $  & $3.6 \pm 0.1 $  \\ 
	  \bottomrule			
		\end{tabular}
		
			\caption{The average $\mathrm{R_{dq}}$ derived by the EMG fit with 95\% confidence interval.}
		\label{tab:Rdq_emg_fit_vendor_mlv}
\end{table}
As seen in the histograms and quantified by the values of the EMG parameters, the cavities produced by RI have on average a smaller $\overline{\mathrm{R_{dq}}}$ of 17\% in the welding seam region in comparison to cavities produced by EZ. The observed vendor specific surface roughness, see Figure \ref{Vendor_Rdq_Distribution} and Table and \ref{tab:Rdq_emg_fit_vendor_mlv}, are in good agreement with the surface topology of EP and BCP treated cavities. It is known that EP leads to a smaller average roughness than BCP \cite{Padamsee2008,Ciovati2008}, as well as smaller boundary step heights and slopes \cite{Geng1999,Xu2011}, which is reflected in the average $\mathrm{R_{dq}}$. 

\subsection{Surface Properties and RF Performance}
\begin{figure}[!htb]
	\centering
		\includegraphics[width=0.9\textwidth]{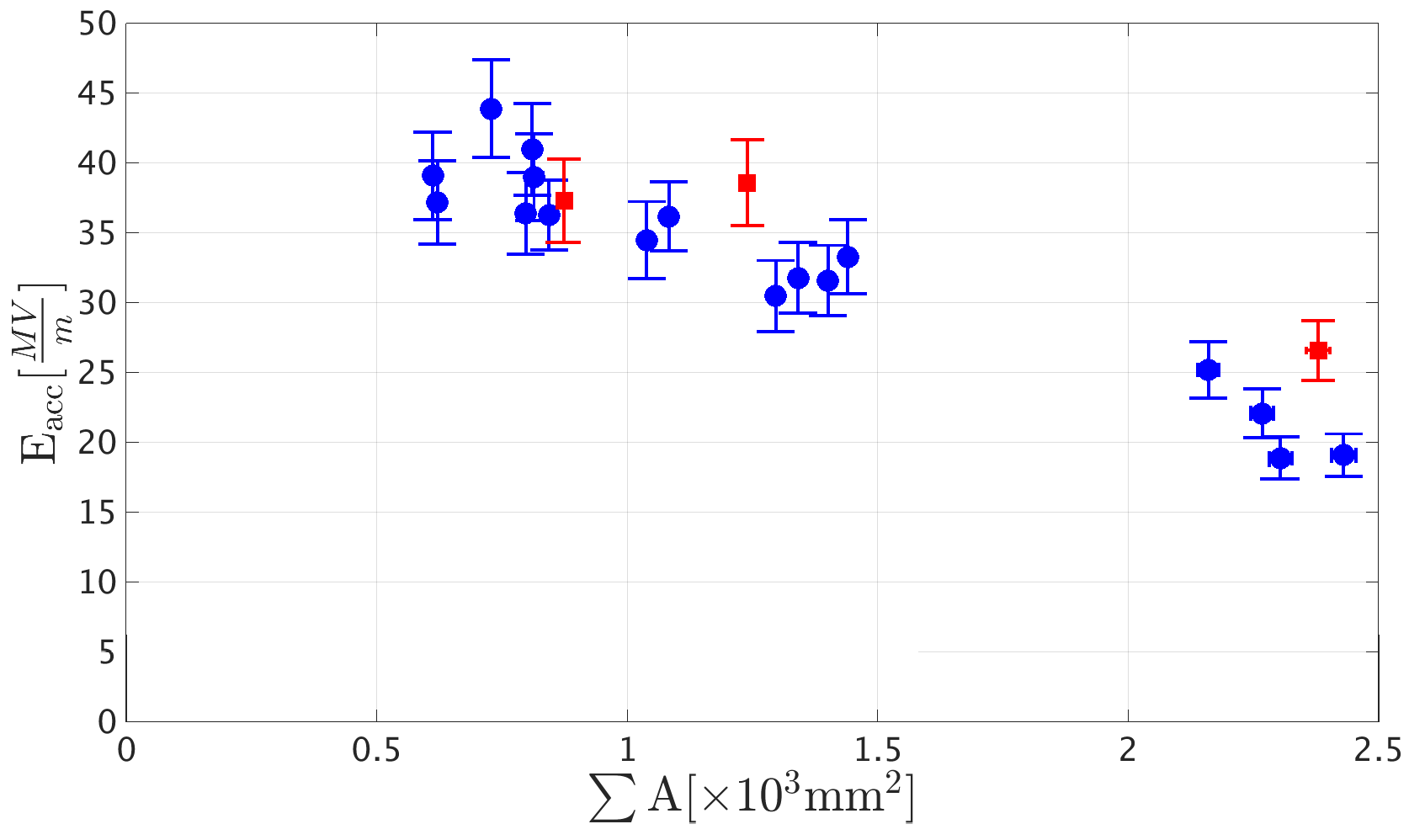}
	\caption{On the x-axis, the integrated grain boundary area of the optically conspicuous cell, which stands for a complete cavity, is shown, while the y-axis depicts the maximum accelerating field achieved by the respective cavity. The red squares display large grain cavities, the blue circles fine grain cavities.}
	\label{fig:Eacc_sumarea_wLG}   
\end{figure}

In order not to be affected by local defects and to deduce an unbiased correlation between optical surface properties and the RF performance of a cavity, a set of cavities are selected by the following criteria:
\begin{enumerate}
	\item No surface chemistry between optical inspection and the cold RF test
	\item Optical inspection shows no local defect
	\item No field emission during the RF test. 
\end{enumerate}
The first criterion is needed to assure that the results of the two methods, optical inspection and cold RF test, can be correlated. A local defect, which is more likely to be the cause of a possible limitation of the cavity RF performance, has to be avoided in order to study the correlation between global surface properties and the RF performance. This is the reason for the second criterion. The last criterion prevents a falsification of the RF performance, because field emissions are a different loss mechanism. A total number of 17 cavities from the XFEL production fulfill the before mentioned criteria, nine RI and eight EZ cavities. To increase the data set, but also to improve the universality of this study, three large grain cavities were included, namely AC151, AC153 and AC154\footnote[1]{The grain size for XFEL cavities had to be on average ASTM 5 or 50\,$\upmu \mathrm{m}$ for the sheets before welding while large grain cavities have grain sizes on the order of several centimeters. These cavities were produced at RI and underwent a main BCP with a total removal of 100-110\,$\upmu \mathrm{m}$.}. They fulfill the above mentioned criteria and increase the data set to a total number of 20 9-cell cavities. The purpose of this investigation is to identify the RF limiting cell. Two assumption were made for this analysis. First, that the maximal accelerating field $\mathrm{E_{acc,max}}$ shows a negative correlation to an optical determined variable. Second, the optically worst cell (aka the cell with the maximum value of this optical variable) should also be the RF limiting cell, since one bad performing cell is sufficient for a cavity to show a bad performance. Hence, the maximum value of the yet to identify optical surface variable of the nine cells identifies the \textit{optically conspicuous cell}. This maximum value is used to represent the whole cavity.
\newline
The different loss models discussed in the literature predict different correlations between the RF performance and the surface properties, but all of them see the grain boundaries as a potential source for limitations or losses \cite{Visentin2003,Bauer2004,Ciovati2007,Ciovati2008a,hylton1988,Safa1999,Knobloch1999a}. Hence the \textit{integrated grain boundary area} $\sum{\mathrm{A}}$ in a cell as optical surface variable is used as a correlator within this work. This property is the sum of all grain boundary areas found in the 75 images of an equator. Since both variables, the grain boundary area and the maximal accelerating field, are subject to measurement uncertainties, the calculation of the correlation coefficient should include these uncertainties. With appropriate estimators, the corrected Pearson correlation coefficient can be deduced \cite{Pearson1966,Darmstadt2011} which includes the influence of the uncertainty on the correlation coefficient.
Figure \ref{fig:Eacc_sumarea_wLG} shows the measured maximum accelerating field as a function of the integrated grain boundary area $\sum{\mathrm{A}}$.
The correlation coefficients are given in Table \ref{tab:Eacc_sum_area}.
\begin{table}[ht]
\small
	\centering
\makebox[\linewidth]{
		\begin{tabular}{l>{\centering\arraybackslash} p{1cm}>{\centering\arraybackslash}    p{1cm}>{\centering\arraybackslash}  p{1cm}>{\centering\arraybackslash}  p{1cm}}
 \toprule
												& All  & RI & E.Z. & Sheets \\
								 \cmidrule{2-5}												
		\textbf{Corrected Pearson $\rho$}  		    &  -0.93 & -0.90 & -0.96 & -0.97    \\      
		\textbf{Significance $\upsigma$}  & 6  & 3.4 & 4 & 5.2 \\
		\bottomrule
		\end{tabular}
		}
		
			\caption{Pearson correlation coefficients for the variable $\sum{\mathrm{A}}$ and $\mathrm{E_{acc,max}}$ for different subgroups.}
		\label{tab:Eacc_sum_area} 
\end{table}
A very strong negative correlation of $\rho$ = -0.93 between the two variables $\sum{\mathrm{A}}$ and $\mathrm{E_{acc,max}}$ was found. The result is consistent for different subgroups, which were tested to reveal systematic origins of a possible correlation. Those three subgroups are the (1) nine RI cavities, (2) the eight EZ cavities and (3) eleven cavities from both vendors but from the same niobium supplier (Sheets). The large grain cavities are only included in the complete sample. The statistical significance of the correlation was found to be $6\upsigma$, the reduced significance for the subgroups is caused by a reduced number of cavities in these datasets.

\section{Conclusion}
The optical inspection robot OBACHT at DESY is well designed and an excellent tool for the automated inspection of superconducting cavities. The time needed for an inspection went down from 2-3 days to half a day. In addition, the reproducibility and accuracy of the images increased as well as the total amount of inspected surface. This goes along with a larger amount of images which need to be reviewed. For the implementation of the optical inspection robot OBACHT as a quality assurance and control tool in a large scale production, as well as a R\&D tool, an automated image processing and analysis algorithm was needed. This algorithm, as presented in this paper, is the first working version of such a code. First results showed that a quality assurance and control is realized. Deviations from the desired cavity properties were detected and the information was given to the cavity vendors. An optical standard for each vendor was identified and in addition, a correlation between the obtained optical surface properties and the cavity performance was observed.

\section{Acknowledgments}
The author would like to thank S. Aderhold (now FNAL), S. Karstensen (DESY), A. Navitski (now RI), J. Schaffran (DESY), L. Steder (DESY) and Y. Tamashevich (now HZB) for their work. Otherwise, mine would have been impossible. This work is funded from the EU 7th Framework Program (FP7/2007-2013) under grant agreement number 283745 (CRISP) and "Construction of New Infrastructures - Preparatory Phase", ILC-HiGrade, contract number 206711, BMBF project  05H12GU9, and Alexander von Humboldt Foundation


\begin{thebibliography}{99}
\bibitem{XFEL_TDR} M. Altarelli et al., {\it The Technical Design Report of the European XFEL} DESY Report 2006-097
\bibitem{ILC_TDR} T. Behnke et al., {\it The International Linear Collider Technical Design Report-Volume 1: Executive Summary}, arXiv:1306.6327
\bibitem{LCLS-II} P. Bishop et al., {\it LCLS-II SRF Cavity Processing Protocol Development and Baseline Cavity Performance Demonstration} in Proceedings of SRF 2015, Whistler, Canada, p. 159-163 
\bibitem{Iwashita2008} Y. Iwashita et al., {\it Development of high resolution camera for observations of superconducting cavities}, Phys. Rev. Accel. Beams 11(9), p. 1-9, 2008 
\bibitem{Tajima2008} T. Tajima et al., {\it Conceptual design of automated systems for SRF cavity optical inspection and assembly} in Proceedings of EPAC 2008, Genova, Italy, p. 922-924
\bibitem{Watanabe} K. Watanabe et al., {\it Review of optical inspection methods and results} in Proceedings of SRF 2009, Berlin, Germany, p. 123-128
\bibitem{Moller2009} W.-D. M\"oller, {\it Review of results from temperature mapping and subsequent cavity inspection} in  Proc. SRF 2009, Berlin, Germany, p. 109-112
\bibitem{Geng2009a} R. L. Geng, G. Ciovati and A. Crawford, {\it Gradient Limiting Defects in 9-Cell Cavities EP Processed and RF at Jefferson Lab},  Proc. SRF 2009, Berlin, Germany, p. 370-374 
\bibitem{Aderhold2010b} S. Aderhold, {\it Optical Inspection of SRF Cavities at DESY Comparison of Optical Inspection and T-Map Data},  Proc. IPAC 2010,  Kyoto, Japan, p. 2896-2898
\bibitem{Singer2010} W. Singer et al., {\it Surface investigation on prototype cavities for the European X-ray Free Electron Laser},  Proc. IPAC 2011, San Sebastian, Spain, p. 2902-2904
\bibitem{Sebastian} S. Aderhold, {\it Study of Field-Limiting Defects in Superconducting RF Cavities for Electron-Accelerator}, Ph.D. Dissertation, Physikalische Fakult\"at, Universit\"at Hamburg, Germany, 2015
\bibitem{Lemke} M. Lemke et al., {\it OBACHT - Optical Bench for Automated Cavity Inspection with High Resolution on Short Time Scales}, ILC-HiGrade Report-2013-01, 2013
\bibitem{Wenskat2015} M. Wenskat, {\it Automated Surface Classification of SRF Cavities for the Investigation of the Influence of Surface Properties onto the Operational Performance}, Ph.D. Dissertation, Physikalische Fakult\"at, Universit\"at Hamburg, Germany, 2015
\bibitem{Iwashita2009} Y. Iwashita et al., {\it R\&D of Nondestructive Inspection Systems for SRF Cavities} in Proceedings of SRF 2009, Berlin, Germany, p. 297-299
\bibitem{Liu2013} X. Liu et al., {\it Single-Image Noise Level Estimation for Blind Denoising},  IEEE Transactions on Image Processing 22(12), p. 5226-5237, 2013 
\bibitem{Sara1995} H. Safa, {\it An Analytical Approach for Calculating the Quench Field in Superconducting Cavities} in Proceedings of SRF 1995, Gif-sur-Yvette, France, p. 413-418
\bibitem{Xie2009} Y. Xie et al., {\it Thermal modeling of ring-type defects} in Proceedings of SRF 2009, Berlin, Germany, p. 331-333
\bibitem{Sage2014} H. Kirschner et al., {\it 3-D PSF Fitting for Fluorescence Microscopy: Implementation and Localization Application},  Journal of Microscopy 249(1), p. 13-25, 2008
\bibitem{Griffa2010} A. Griffa et al., {\it Comparison of Deconvolution Software in 3D Microscopy: A User Point of View},  G.I.T. Imaging \& Microscopy 12(1), p. 43-45, 2010
\bibitem{Born2003} M. Born, {\it Fraunhofer Diffraction at Apertures of Various Forms},  in Principles of optics, 7th edition, Cambridge University Press, 2003, Chapter 8.5, pages 436-445.
\bibitem{Frisken-Gibson1989} S. Frisken-Gibson and F. Lanni, {\it Diffraction by a circular aperture as a model for three-dimensional optical microscopy},  JOSA A 6(9), p. 1357-1367, 1989
\bibitem{FS1} E. Adelson et al., {\it Pyramid methods in image processing}, RCA Engineer 29(6), p. 33-41, 1984
\bibitem{Forster2004} B. Forster et al., {\it Complex wavelets for extended depth-of-field: A new method for the fusion of multichannel microscopy images}, Microscopy research and technique, 65(1‐2), p. 33-42, 2004
\bibitem{Gresele2013} A. Gresele, {\it The Statistics of industrial XFEL Cavities fabrication at E. Zanon} in Proceedings of SRF 2013, Paris, France, p. 180-182
\bibitem{Sulimov} A. Sulimov, {\it The Statistics of industrial XFEL Cavities fabrication at Research Instruments} in Proceedings of SRF 2013, Paris, France, p. 9-11
\bibitem{Raquel} R. Gomez-Ambrosio, {\it Autofocus System Implementation for Optical Inspection of Superconducting Cavities}, Summerstudent Report DESY 2010
\bibitem{Autofocus1} A. Santos et al., {\it Evaluation of Autofocus Functions in Molecular Cytogenetic Analysis},  Journal of microscopy 188(3), p. 264-272, 1997
\bibitem{Autofocus2} Y. Sun et al., {\it Autofocusing in Computer Microscopy: Selecting the Optimal Focus Algorithm}, Microscopy research and technique 65(3), p. 139-149, 2004
\bibitem{Toropov} E. Toropov and D. Sergatskov, {\it Automation of Optical Inspection in Fermilab}, Technical Report Fermilab TD-11-004, 2011
\bibitem{Steel} M. Lemke et al., {\it  Methods and optimisation for OBACHT lighting}, ILC-HiGrade Report-2015-02, 2015
\bibitem{Watanabe2008} K. Watanabe, {\it Recent inspection results by Kyoto-camera} in Proceedings of of TTC2008, New Delhi, India
\bibitem{Martin2004} D. Martin, C.C. Fowles and J. Malik, {\it Learning to detect natural image boundaries using local brightness, color and texture cues},  IEEE transactions on pattern analysis and machine intelligence 26(5), p. 530-549, 2004
\bibitem{COL:COL20020} A. D. Broadbent, {\it A critical review of the development of the CIE1931 RGB color-matching functions},  Color Research \& Application 29(4), p. 267-272, 2004
\bibitem{Pizer1987} S. M. Pizer, E. P. Amburn and J. D. Austin, {\it Adaptive Histogram Equalization and Its Variations}, Computer Vision, Graphics and Image Processing 39, p. 355-368, 1987
\bibitem{Zuiderveld1994} K. Zuiderveld, {\it Contrast limited adaptive histogram equalization}, Graphics gems IV (Academic Press Professional, Inc. San Diego, CA,), p. 474-485, 1994 
\bibitem{Arias-Castro2009} E. Arias-Castro, D. L. Donoho, {\it Does median filtering truly preserve edges better than linear filtering?}, The Annals of Statistics 37(3), p. 1172-1206, 2009
\bibitem{Otsu1979} N. Otsu, {\it A Threshold Selection Method from Gray-Level Histograms}, IEEE Transactions on Systems, Man and Cybernetics 9, p. 62-66, 1979
\bibitem{Cherry67} A. H. Robinson et al. . {\it Results of a prototype television bandwidth compression scheme}, Proceedings of the IEEE 55 (3), p. 356–364, 1967
\bibitem{dulmage1958} A. L. Dulmage and N. S. Mendelsohn, {\it Coverings of bipartite graphs}, Canadian Journal of Mathematics 10(4), p. 516-534, 1958 
\bibitem{Hall01011935} P. Hall, {\it On Representatives of Subsets},  Journal of the London Mathematical Society 10(1), p. 26-30, 1935
\bibitem{Jaehne2004} B. J\"ahne, {\it Practical Handbook on Image Processing for Scientific and Technical Applications}, CRC Press (1997), Figure 10.23a, page 348
\bibitem{Hu1962} M. Hu, {\it Visual pattern recognition by moment invariants},  IRE Transactions on Information Theory 8(2), p. 179-187, 1962
\bibitem{Patil2011} S. B. Patil and K. B. Shrikant, {\it Image Processing Method To Measure Sugarcane Leaf Area},  International Journal of Engineering Science and Technology (3)8, p. 6394-6400, 2011
\bibitem{Klette1999} R. Klette and J. Zunic, {\it On errors in calculated moments of convex sets using digital images} in Proceedings of SPIE 3811, Vision Geometry VIII 105, p. 82-94, 1999
\bibitem{Liao1993} S. Laio, {\it Image Analysis by Moments}, Ph.D. Dissertation, Department of Electrical and Computer Engineering, University of Manitoba, 1993
\bibitem{iso25178} International Organization for Standardization 2009, {\it ISO 25178 - Geometric Product Specifications (GPS) Surface texture: areal}
\bibitem{young1998} I. T. Young, J. J. Gerbrands and L. J. van Vliet, {\it Fundamentals of Image Processing}, Delft University of Technology (1998), Chapter 6 - Noise, 1998
\bibitem{Singer2013} W. Singer et al., {\it Development of large grain cavities}, Phys. Rev. Accel. Beams 16(1), p. 12003, 2013
\bibitem{Knobloch1999a} J. Knobloch et al., {\it High-field Q slope in superconducting cavities due to magnetic field enhancement at grain boundaries} in Proceedings of SRF 1999, Santa Fe, USA, p. 77-91
\bibitem{CORDIS} E. Elsen (2015, Jan.). ILC-HiGrade Report Summary [Online] Available: $http//cordis.europa.eu/result/rcn/156016\_en.html$
\bibitem{Singer2016} W. Singer et al., {\it Production of superconducting 1.3-GHz cavities for the European X-ray Free Electron Laser},  Phys. Rev. Accel. Beams 19(9), p. 092001, 2016  
\bibitem{Wenskat2015a} M. Wenskat and L. Steder, {\it Characterization of optical surface properties of 1.3 GHz cavities for the European XFEL} in Proceedings of SRF 2015, Whistler, Canada, p. 795-798
\bibitem{Yegor2017} Y. Tamashevich, {\it Diagnostics and treatment of 1.3 GHz Nb Cavities}, Ph.D. Dissertation, Physikalische Fakult\"at, Universit\"at Hamburg, Germany, 2017
\bibitem{NavitskiSRF2015} A. Navitski et al., {\it Characterization of Surface Defects on EXFEL Series and ILC-HiGrade Cavities} in Proceedings of SRF 2015, Whistler, Canada, p. 281-285
\bibitem{Aderhold2010} S. Aderhold et al., {\it Cavity Process}, ILC-HiGrade Report-2010-05, 2010 
\bibitem{EBW1997} S. Lampman, {\it Weld Integrity and Performance}, ASM International (1997), Chapter 1 - Weld Solidification
\bibitem{Udomphol2007} T. Udomphol, {\it Weld Microstructure}, Suranaree University of Technology (2007) 
\bibitem{EBWBuch} Schultz,H 2000 {\it Elektronenstrahlschweissen 2nd Edition} (DVS-Verlag GmbH, D\"usseldorf), p. 41
\bibitem{Singer2017} W. Singer, {\it Fabrication of elliptical SRF cavities}, Supercond. Sci. Technol. 30, p. 033001, 2017
\bibitem{Geng1999} R. L. Geng, J. Knobloch and H. Padamsee, {\it Mirco-Structures of RF Surfaces in the Electron-Beam-Weld Regions of Niobium} in Proceedings of SRF 1999, Santa Fe, USA, p. 238-245
\bibitem{Schmidt10} Schmidt, A et al. 2010 "1.3 GHz Niobium Single-Cell Fabrication Sequence" TTC-Report-2010-01
\bibitem{Padamsee2008} H. Padamsee, J. Knobloch and T. Hays, {\it RF Superconductivity for Accelerators}, Weinheim: Wiley-VCH Verlag (2008), 
\bibitem{Ciovati2008} G. Ciovati et al., {\it Final Surface Preparation for Superconducting Cavities}, TTC-Report-2008-05, 2008
\bibitem{Xu2011} C. Xu et al., {\it Enhanced Characterization of Niobium Surface Topography}, Phys. Rev. Accel. Beams 14, p. 123501, 2011
\bibitem{Visentin2003} B. Visentin, {\it Q-Slope at High Gradients: Review about Experiments and Explanations} in Proceedings of SRF 2003, L\"ubeck, Germany, p. 199-205
\bibitem{Bauer2004} P. Bauer, {\it Review of Models of RF Surface Resistance in High Gradient Niobium Cavities for Particle Accelerators}, Technical Report TD-04-014, 2004
\bibitem{Ciovati2007} G. Ciovati and A. Gurevich, {\it Measurement of RF losses due to trapped flux in a large grain niobium cavity} in Proceedings of SRF 2007, Beijing, China, p. 132-136
\bibitem{Ciovati2008a} G. Ciovati and A. Gurevich, {\it Evidence of high-field radio-frequency hot spots due to trapped vortices in niobium cavities}, Phys. Rev. Accel. Beams 11, p. 1-12, 2008
\bibitem{hylton1988} T. Hylton et al., {\it Weakly coupled grain model of high-frequency losses in high Tc superconducting thin films}, ApPhL 14, p. 1343-1345, 1988
\bibitem{Safa1999} H. Safa, {\it Specific Resistance Measurement of a Single Grain Boundary in Pure Niobium} in Proceedings of SRF 1999, Santa Fe, USA, p. 267-269
\bibitem{Pearson1966} W. Pearson, {\it Estimation of a correlation coefficient from an uncertainty measure}, Psychometrika 31(3), p. 421-433, 1966
\bibitem{Darmstadt2011} TU Darmstadt (2017, Feb.), {\it Berechnung der Korrelation bei Vorliegen von Messfehlern} [Online] Available: $http://www.zfs.tu-darmstadt.de/media/zfs/materialien\_4/8\_Korrleation.pdf$ 
\end{thebibliography}
\end{document}